\documentclass[fleqn,usenatbib]{mnras}
\usepackage{newtxtext,newtxmath}
\usepackage[T1]{fontenc}

\DeclareRobustCommand{\VAN}[3]{#2}
\let\VANthebibliography\thebibliography
\def\thebibliography{\DeclareRobustCommand{\VAN}[3]{##3}\VANthebibliography}

\usepackage{graphicx}	
\usepackage{amsmath}	
\usepackage{bm}         
\usepackage{multicol}   
\usepackage{pdflscape}	
\usepackage{xcolor}	    
\usepackage{orcidlink}
\usepackage{subcaption}
\usepackage{makecell}
\usepackage{footnote}
\usepackage{threeparttable}

\newcommand{\kms}{\,km\,s$^{-1}$}	

\title[dC stars are a halo population]{Carbon-enhanced dwarf stars are predominantly a halo population}

\author[J.~Farihi et al.]{Jay Farihi,$^{1}$\thanks{E-mail: j.farihi@ucl.ac.uk}
Jason L.~Sanders,$^{1}$
Sophia Lilleengen,$^{2,3}$
Lewis J.~Whitehouse,$^{1}$
and Denis Erkal$^{3}$
\\
$^{1}$Department of Physics and Astronomy, University College London, London WC1E 6BT, UK\\
$^{2}$Institute for Computational Cosmology, Department of Physics, Durham University, Durham DH1 3LE, UK \\
$^{3}$School of Mathematics and Physics, University of Surrey, Guildford, GU2 7XH, UK\\
}

\date{Accepted XXX. Received YYY; in original form ZZZ}

\pubyear{2025}

\begin{document}
\label{firstpage}
\pagerange{\pageref{firstpage}--\pageref{lastpage}}
\maketitle

\begin{abstract}
This paper reports a Galactic kinematical and dynamical analysis of 1003 main-sequence carbon stars. The sample is drawn from the Sloan Digital Sky Survey, and cross-matched with {\em Gaia} DR3 to obtain 6-dimensional positions and velocities using a Bayesian framework. The study provides the first reliable distances for a large sample of dwarf carbon stars, which are then analyzed using both space motions and actions. The results are combined with dynamical equilibrium models for the three primary Galactic components to assign membership, finding that around 60\,per cent belong to the halo, and over 30\,per cent originate in the thick disc. Therefore, the results indicate dwarf carbon stars are dominated by a metal-poor halo population, and are thus an excellent resource for stellar archaeology. These stars remain on the main sequence and are relatively nearby, but atmospheric modelling is challenged by their cool effective temperatures and strong molecular features.  In light of this, efforts should be made to improve C/O $>1$ atmospheric modelling, as the subset of low-mass dwarf carbon stars may numerically dominate the Galactic population of carbon-enriched, metal-poor stars.
\end{abstract}




\section{Introduction}

Ancient, metal-poor stars that were born in the early Galaxy and Universe provide important constraints on the formation and evolution of the first stars, and constrain Galactic chemical enrichment over time \citep[see][and references therein]{Beers2005,Frebel2015} that eventually produced the necessary ingredients for the solar system \citep[e.g.][]{Jura2013,Young2014}.  One of the prime challenges in the study of such ancient stars is the identification of their metal poverty, which is gradually being addressed through large surveys using spectroscopy \citep{Christlieb2002,Keller2007}, as well as photometry \citep{York2000,Starkenburg2014}.

Among the myriad classifications for metal-poor stars, and their corresponding metallicities that span several decades, there is a well-established trend of carbon enrichment \citep[e.g.][]{Lucatello2005,Yong2013,Arentsen2022}.  While the origin of these carbon-enhanced metal-poor (CEMP) stars remains an outstanding problem in stellar evolution \citep{Frebel2015,Yoon2016}, there are essentially two possibilities: mass transfer from a carbon-rich, evolved stellar companion \citep[e.g.][]{Abate2015}, or carbon within natal clouds from which the stars formed \citep[e.g.][]{Cooke2011}.  As the number of metal-poor stars identified through continuing observational efforts increases, so too will the number of stars with carbon enhancement.

Dwarf carbon (dC) stars are late-type main-sequence stars with C/O $>1$, with spectra often similar to carbon-rich, asymptotic giant branch stars \citep{Dahn1977}.  Because carbon is synthesized within the cores of solar- and low-mass stars during their main-sequence lifetime, and only later brought to the stellar surface during the third dredge-up \citep{Iben1965}, the occurrence of these stars cannot readily be accounted for with standard, single-star evolution theories.  Intriguingly, the first three confirmed carbon dwarfs have unambiguous halo kinematics and Hertzsprung-Russell (HR) diagram positions well below the main sequence \citep{Harris1998}.  Among these visitors from the halo, the prototype dC star, G77-61, is currently at 73.1\,pc distance \citep[$\varpi=13.7\pm0.1$\,mas;][]{GaiaCollaboration2023}, but is extremely metal-poor ([Fe/H]\,$=-4.0$; \citealt{Plez2005}), suggesting a chemo-dynamical connection to the known CEMP stars located within the Galactic halo.  Thus, at least some fraction of dC stars are CEMP stars by definition.

There are approximately 1500 confirmed and candidate dC stars, mostly identified without kinematical bias through large spectroscopic surveys \citep{Green2013,Li2024}.  Although these are sizable samples, they essentially lack characterization such as masses, effective temperatures, metallicities, and binary fractions, with only a handful of exceptions such as the dC stars detected and monitored astrometrically by the US Naval Observatory for over two decades \citep{Harris2018}. For example, only the bright ($g,r)=(14.6,13.3)$\,AB\,mag, prototype G77-61 has a published $T_{\rm eff}$ and abundances for elements such as C, N, Na, Mg, and Fe \citep{Gass1988}; determinations that are challenging in the presence of strong molecular features, even where high-resolution spectroscopy of these faint stars is feasible \citep{Plez2005}.  While there are no mass estimates for any dC stars as yet, based on the existing (and possibly biased) samples, their positions in the HR diagram suggest temperatures within or near the range 4000--4500\,K, and thus masses below solar \citep[G77-61 is estimated at 0.3\,M$_\odot$;][]{Dearborn1986}.  

In the study of carbon-enhanced stars for stellar archaeology, the dC stars have a major advantage in that late-type dwarfs should be numerically dominant in the Galaxy, based on the initial mass function \citep{Salpeter1955,Reid2002,Kirkpatrick2024}. Furthermore, stars remaining on the main sequence are not subject to any possible mixing scenarios, such as those postulated for CEMP stars \citep{Fujimoto2000}, which are typically on the first ascent giant branch in order to be detected within the distant halo.  However, the larger dC population is not yet amenable to spectroscopic abundances studies, where a typical star has $g>17$\,AB\,mag and $H>13$\,mag, and thus past the magnitude limits of APOGEE in the near-infrared \citep[$H=12.2$\,mag;][]{Majewski2017}.  Furthermore, while their $T_{\rm eff}\lesssim4500$\,K results in the characteristic, strong molecular features of carbon, available atmospheric models are typically carbon-normal at these low temperatures \citep[e.g.][although see \citealt{Aringer2016}]{Gustafsson2008}.

Because reliable and robust atmospheric and chemical modelling is not yet possible for dC stars as a class, an alternative approach to understand their origins is to study their kinematics and Galactic orbits.  A previous study based on {\em Gaia} DR1 proper-motion catalogues of several hundred dC stars, and thus lacking reliable distances and instead using absolute magnitude estimates, suggested the population contains a significant halo component of at least 30\,per cent, and that the majority are kinematically old, likely metal-poor stars in the thick disc and halo \citep{Farihi2018}.  These findings for large numbers of non-kinematically selected stars are consistent with early indications based on a small number of stars detected via high proper motions \citep{Harris1998}. 

With {\em Gaia} DR3 it is now possible to estimate distances to $N\sim1000$ dC stars and thus obtain full 6-dimensional space positions and velocities.  From these, one can calculate Galactic orbits and dynamics.  This paper is organized as follows.  Section 2 discusses the sample of dC star candidates used in this study and their associated data.  Section 3 covers the inferences of distances to these stars using a Bayesian framework, and identifies the subset that can be confidently assigned to the main sequence.  Section 4 details the kinematic and action-based analysis of the high-confidence dC stars, and their assignment to one of the three major components of the Galaxy.  A brief discussion of the findings and future developments is given in Section 5.

\section{Candidate \texorpdfstring{\lowercase{d}C}{dC} stars and data}
\label{sec:data}

Candidate dC stars are sourced from a published catalogue, assembled via cross-correlation between all spectroscopic targets observed during Sloan Digital Sky Survey \citep[SDSS;][]{York2000} DR7 and DR8, and a series of carbon star spectral templates, yielding a total of 1211 unique objects \citep[][table~1]{Green2013}.  While this identification method is biased toward the specific templates utilized to guide the colour selection criteria, it critically provides a kinematically unbiased sample of stars, which have only been selected for spectroscopic fibres based on their colours.  However, because no kinematical cuts were made for the catalogue, it likely contains stars that do not lie on the main sequence, and hence are not dC stars.  The removal of such catalogue contaminants is addressed in the next Section.

For each candidate dC star, all primary and ancillary data are sourced from SDSS DR16 \citep{SDSS_DR16}, and \textit{Gaia} DR3 \citep{GaiaCollaboration2023}.  SDSS $gri$ PSF magnitudes are retrieved using \textsc{astroquery} \citep{astroquery} by cross-matching the catalogue of candidate dC stars with the SDSS database using the J2000 coordinates of each target. With the {\sc bestobjid} and {\sc specobjid} for each source, radial velocities are adopted as those determined by the SEGUE survey and sourced via {\sc sppparams} \citep{Yanny2009}. Parallaxes, $\varpi$, and proper motions, $(\mu_\upalpha,\mu_\updelta)$ are taken from \textit{Gaia} DR3, by performing a cone search around each dC source using a radius of 10\arcsec after propagating the {\it Gaia} astrometry from the reported DR3 epoch J2016.0 back to J2000, the approximate epoch of the SDSS astrometry.

From the initial 1211 candidate dC stars, 1372 matches are returned from the cone search. When more than one {\em Gaia} source is returned for a given SDSS target, the closest on-sky match is taken and checked by comparing the SDSS $r$ brightness to a synthetic $r$-band magnitude computed from the \textit{Gaia} $G$-band mean magnitude and $G_{\rm BP}-G_{\rm RP}$ colour \citep{Riello2021}.  This yielded appropriate matches, except for the star labelled as SDSS\,J111324.84+220916.8 by \citet{Green2013}, which should be matched to the photometric source SDSS\,J111324.94+220916.2 (although this star does not enter the final high-confidence dC catalogue). A total of three candidate dC stars lacked counterparts in the \textit{Gaia} DR3 cross-match, reducing the 1211 candidates to 1208.  All {\it Gaia} parallaxes are zero-point corrected \citep{Lindegren2021}\footnote{\url{https://pypi.org/project/gaiadr3-zeropoint/}}. 

\section{Probabilistic distance inference}\label{sec:distances}

In order to carry out a Galactic dynamical analysis of the dC star sample, full and reliable 6-dimensional position and velocity data are necessary.  It is well known that simply inverting \textit{Gaia} parallax measurements to obtain distances is prone to systematic errors and biases \citep{Luri2018, Lindegren2021}.  An obvious example of these biases is negative parallax (e.g.\ for faint stars in crowded fields) that yields an unphysical distance if inverted. 

Fortunately, there are Bayesian approaches to infer the distances that have been demonstrated to be robust, and which are adopted for the analysis here \citep{Bailer-Jones2015,Astraatmadja2016}.  This inference-based methodology has produced a distance catalogue for all stars that have a parallax in {\em Gaia} DR3 \citep{Bailer-Jones2021}, including the dC star candidates in this study.  A Bayesian approach produces chains (posterior samples) of values for distances for each star based on its observational uncertainties.  However, the \citet{Bailer-Jones2021} distance catalogue reports the distances and errors as the median, $16^{\rm th}$, and $84^{\rm th}$ percentiles of these chains but does not provide the chains themselves.  Distances need to be drawn from a distribution to perform a statistical analysis of the dC star sample, and it is not possible without the chains and shape of the distance distribution for each star.  

Therefore, the `metric' method is adopted to infer the distances for the sample of dC stars. The observational data from the {\it Gaia} DR3 astrometry and SDSS photometry are jointly modeled as $\bm{y}=[\varpi, \mu_\upalpha, \mu_\updelta, g, r, i]$. with associated uncertainty covariance matrix $\upsigma_{\bm{y}}$ (as provided in the {\it Gaia} DR3 catalogue and assuming photometric errors are uncorrelated). For each star, the goal is to infer the distribution of a set of model parameters $\bm{\eta}=[d, \mu_\upalpha', \mu_\updelta', g_\mathrm{mod}, r_\mathrm{mod}, i_\mathrm{mod}]$ where $d$ is the distance, $(\mu_\upalpha', \mu_\updelta')$ the true proper motion, and $(g,r,i)_\mathrm{mod}$ the (extinction-corrected) model magnitudes. The posterior is given by
\begin{equation}
    p(\bm{\eta}|\bm{y}) \propto \mathcal{L}\left(\bm{y}|\bm{\eta}\right) p_\mathrm{Gaia}(d, \mu_\upalpha', \mu_\updelta') p_\mathrm{CMD}(g,r,i|d),\label{eqn::posterior}
\end{equation}
where the first term is the likelihood, the second term is an astrometric prior, and the third term is a photometric prior. The likelihood is computed as 
\begin{equation}
    \log\mathcal{L}\left(\bm{y}|\bm{\eta}\right) \propto -\frac{1}{2} \left(\bm{y}-f(\bm{\eta})\right)^\mathrm{T} (\upsigma_{\bm{y}}+\upsigma_{f(\bm{\eta})})^{-1} \left(\bm{y}-f(\bm{\eta})\right),
\end{equation}
where the function $f(\bm{\eta}) = (1/d, \mu_\upalpha', \mu_\updelta', g+A_g(d), r+A_r(d), i+A_i(d))$, and $A_x(d)$ is the extinction in the $x$ band at distance $d$. Extinction corrections are computed using 3-dimensional extinction maps \citep[using the {\sc dustmaps} interface,][]{Green2018,Green2019_ext}. These are converted to extinction in the $(g,r,i)$ bandpasses using the coefficients $(3.651,2.633,2.046)$, which are estimated by scaling the reported Pan-STARRS $r$-band coefficient by the ratio of the SDSS to Pan-STARRS $r$-band coefficients, and then ensuring the SDSS colour excess ratios follow that of Pan-STARRS \citep{SchlaflyFinkbeiner}. \citet{Green2019_ext} provide uncertainties in the extinction at each distance, which are incorporated into the likelihood by propagating the resulting (correlated) uncertainty in $f(\bm{\eta})$, given by the covariance matrix $\upsigma_{f(\bm{\eta})}$, whose astrometric components are all zero.

Two prior contributions are introduced in Equation~\ref{eqn::posterior}, $p_\mathrm{Gaia}$ and $p_\mathrm{CMD}$, that define the prior on the {\it Gaia} astrometry and the prior on the photometry respectively.  Each of these priors is now defined.

\subsection{Astrometric prior}
The astrometric prior is split into two terms: $p_\mathrm{Gaia}(d, \mu_\upalpha, \mu_\updelta)=p(d)p(\mu_\upalpha, \mu_\updelta | d)$, following \citet{Bailer-Jones2021} and using a gamma distribution for the distance prior
\begin{equation}
    p(d)\propto
    \begin{cases}
        d^{\upbeta} e^{-(d/L)^{\upalpha}}   & \text{if } d\geq 0,\\
        0                           & \text{otherwise.}
    \end{cases}
\end{equation}
This distance prior generically takes into account the exponentially decreasing space density of stars beyond some length scale $L$ \citep{Bailer-Jones2015}, although the gamma distribution allows for flexibility around this when looking at stars nearer the Galactic midplane, for example. \citet{Bailer-Jones2021} provide appropriate choices for the $p(d)$ parameters $(L,\upalpha,\upbeta)$, with on-sky position based on fits to the {\it Gaia} mock simulation, GeDR3mock. These parameter choices are utilised in the following.

In addition to information from the parallax, there is weak information on the distance from the proper motions. If it is assumed that a typical star is bound to the Galaxy, the distance cannot be so large that, for a fixed proper motion, the star would be unbound. This is imposed by requiring stars have a maximum, heliocentric tangential velocity of 750\kms \citep{Bailer-Jones2018, Luri2018}, where $v_\mathrm{tan}=4.74 d \sqrt{\mu_\upalpha^2+\mu_\updelta^2}$  such that $p(\mu_\upalpha, \mu_\updelta | d) \propto d\,p(v_\mathrm{tan})$ and

\begin{equation}
    p(v_\mathrm{tan}) = 
    \begin{cases}
        \frac{1}{B\left(a, b\right)} 
        \left(\frac{v_\mathrm{tan}}{v_\mathrm{max}}\right)^{a-1}
        \left(1-\frac{v_\mathrm{tan}}{v_\mathrm{max}}\right)^{b-1}
        & \text{if } 0\leq v_\mathrm{tan}\leq v_\mathrm{max},\\
        0              & \text{otherwise.}
    \end{cases}
\end{equation} 
This is a beta function with the parameters $a=2$, $b=3$ that control the shape of the distribution. This prior favours uniformly distributed velocities at small velocity and the high velocity taper prevents the possibility of arbitrarily large velocity solutions.  No priors are placed on the direction of the tangential velocity.

\subsection{Photometric prior}

Colour (absolute) magnitude diagrams (CMDs) are a tool analogous to the HR diagram for visualising stellar properties and evolution, using colour in place of effective temperature, and absolute magnitude as a proxy for luminosity.  The structure of a CMD implies that for any given colour, a non-uniform probability distribution of absolute magnitude exists. Furthermore, through the distance modulus, the absolute magnitude contains information on the true distance of any given star. Thus, prior constraints can be placed on the true distance of a target based on its colour \citep{Bailer-Jones2011}. This is encoded in the CMD prior, $p_\mathrm{CMD}(g,r,i | d)$, the prior probability of finding a star with given (error-free, de-reddened) SDSS photometry given its distance. It should be noted that $p(g,r,i | d)$ is equivalent to $p(M_r, g-r, r-i)$, the probability of a given extinction-corrected absolute $r$-band magnitude and the two de-reddened colours, $(g-r)$ and $(r-i)$.

The first step toward constructing such a prior is to assemble a high-fidelity CMD. Because the candidate dC stars are sourced from the SDSS, and hence possess multi-band photometry, the CMD priors are constructed using stars with excellent astrometry from \textit{Gaia} DR3, and photometry from SDSS DR16.  A set of 250\,000 stars is generated by randomly querying the \textit{Gaia} DR3 cross-match with the SDSS DR13 best neighbour catalogue\footnote{The most recent data release to possess a cross-match provided by the \textit{Gaia} team.} All stars returned by this query are required to have re-normalized unit weight error below 1.2 \citep{Belokurov2020}, and a measured parallax signal-to-noise above 50.  Distances to each of the 250\,000 stars are taken from the \textit{Gaia} photo-geometric distance catalogue \citep{Bailer-Jones2021}, but the inverted parallaxes are likely also reliable given the quality cuts.

Photometric measurements in the SDSS $gri$ bands for each of the stars with the aforementioned quality cuts are retrieved by querying the DR16 database using the {\sc bestobjid} in {\sc casjobs}. The photometric quality flags for each star in the pool are inspected to ensure that all targets possess good-quality photometric measurements. Extinction corrections are computed using the median of the aforementioned 3-dimensional extinction maps \citep{Green2018,Green2019_ext}, evaluated at the median \textit{Gaia} photo-geometric distance (using the coefficients described previously). Extinction-corrected SDSS $r$-band absolute magnitudes, $M_r$, and the extinction-corrected colours $(g-r)$ and $(r-i)$ are computed for the 250\,000 stars to assemble CMDs. SDSS $u$-band photometry is omitted because dC stars are faint in this band owing to their cool effective temperatures, exacerbated by strong carbon molecular absorption. Additionally, because of difficulties with ground-based photometry in the red (e.g.\ a less efficient detector, and telluric features), SDSS $z$-band photometry is also omitted \citep{Doi2010}. We compute the prior, $p(M_r,g-r,r-i)$, from the density of the 250\,000 reference stars using a kernel density estimate evaluated by fast Fourier transform with the \textsc{KDEpy} package\footnote{\url{https://github.com/tommyod/KDEpy}}. A Gaussian kernel with a standard deviation of $0.1$ is adopted. For speed, the density estimate is evaluated on a fine grid of $200$ points in each dimension with $-2$\,AB\,mag $\le M_{r} \le 16$ AB\,mag, $-0.5 \le (g-r) \le 2.5$, and $-0.5 \le g-i \le 2.0$, which can then be interpolated.

It is important to note that these priors may introduce bias in distance determination, as they pull all posterior samples toward the main sequence. dC stars naturally have different colours to carbon-normal main-sequence stars. Furthermore, there is a handful of dC stars known to exhibit a composite spectrum with a white dwarf, but these represent less than 1\,per cent of the sample \citep{Green2013}. However, the candidate dC stars are faint with $\langle r \rangle \approx 18.5$\,AB\,mag, and have modest- to poor-quality parallax measurements; thus, the use of CMD priors is strongly beneficial to the distance inference.

\begin{figure}
    \includegraphics[width=\columnwidth]{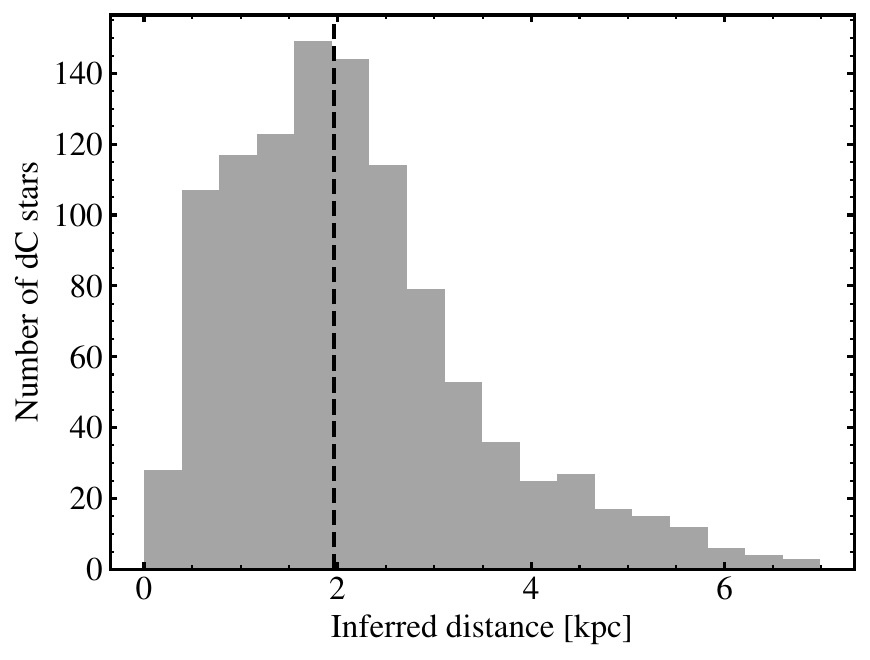}
    \caption{Distribution of the resulting median distance for 1003 high-confidence dC stars, inferred from the distance modulus based on the results from the CMD diagram analysis.  The black dashed line marks the sample median distance at $d=1.96$\,kpc, and the bin width equals the median of the sample distance uncertainty.}
    \label{fig:distances}
\end{figure}

\begin{figure}
    \includegraphics[width=\columnwidth]{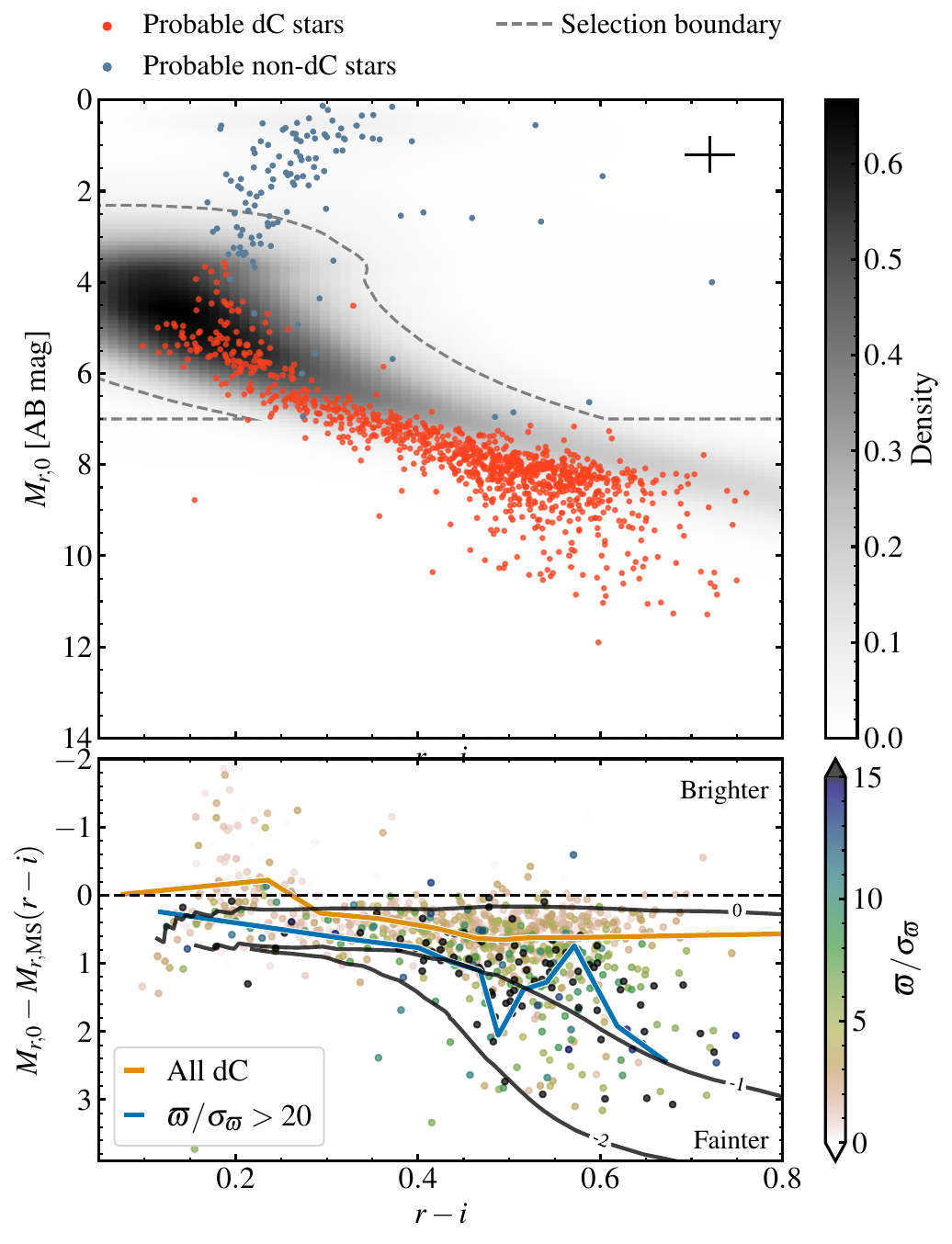}
    \caption{Colour-absolute magnitude diagram of the 1208 candidate dC stars. \textit{Top}: The $x$-axis is the de-reddened SDSS $(r-i)$ colour, and the $y$-axis is the de-reddened absolute magnitude in the SDSS $r$ band, calculated from the inferred distance.  The greyscale shading, indicated by the colour bar on the right, is the underlying prior density distribution.  The dashed grey line shows the boundary below which stars are considered high-confidence dC stars, defined as where the prior density is above 0.08 or $M_r>7$\,AB mag. The dark orange circles mark the candidates selected as high-confidence dC stars, and the blue circles are rejected (those within the selection boundary here are excluded based on CMDs using $(g-r)$ or $(g-i)$; Figure~\ref{fig:extra_cmd}). The error bar in the top right corner shows the median uncertainties in $(r-i)$ and $M_r$ (using $16^{\rm th}$ and $84^{\rm th}$ percentiles) for all candidates.  This selection procedure results in 1003 high-confidence dC stars. \textit{Bottom}: Difference between the de-reddened $r$-band absolute magnitude and the peak prior density at each $(r-i)$, coloured by the parallax signal-to-noise, $\varpi/\upsigma_\varpi$. The orange line shows the median for the high-confidence dC stars and the blue the subset with $\varpi/\upsigma_\varpi>20$. The three black lines are model isochrones \citep{Dotter2008} with metallicities $[\mathrm{Fe/H}]=0,-1,-2$ as labelled.}
    \label{fig:CMD}
\end{figure}

\subsection{Resulting photo-geometric distances for dC stars}\label{sec:dist:res}

The posterior of the model parameters given the data (Equation~\ref{eqn::posterior}) are sampled using the MCMC package {\sc emcee} \citep{Foreman-Mackey2013}. For each dC candidate, an MCMC run starts with 50 walkers, each taking 5000 steps. The walkers are initialized by randomly drawing from a normal distribution centred at the measurement value for $(\mu_\upalpha, \mu_\updelta, g,r,i)$, and with a standard deviation equal to the measurement uncertainty. The initialization of the distances is done similarly, but with a mean $1/\varpi$ if $\varpi>0.2\,\mathrm{mas}$, otherwise $5\,\mathrm{kpc}$ and a standard deviation of $30$\,per cent.  All chains are deemed to have converged from the calculation of the autocorrelation times.  The initial 1500 steps taken by each walker are discarded, and the median of each sampling chain is taken as the inferred parameter value, with the final error quoted as the $16^{\rm th}$ and $84^{\rm th}$ percentile confidence intervals (owing to the non-Gaussian nature of some posterior probability distributions).  This results in distances for the dC candidates and produces chains that allow for random sampling in subsequent analysis (Section~\ref{sec:class:method}).

The distances inferred from the photo-geometric method are shown in Figure~\ref{fig:distances}, where half are located within 1.96\,kpc, and the remaining stars have distances up to 6.5\,kpc.  These represent the first reliable distances calculated for a statistically compelling sample of $N\sim1000$ dC stars.

With the distances shown in Figure~\ref{fig:distances}, the dC stars are plotted on (de-reddened) colour-absolute magnitude diagrams for each of $M_r$ vs.\ $(g-r)$, $(g-i)$, and $(r-i)$. The $(r-i)$ CMD is shown in Figure~\ref{fig:CMD}, and the other CMDs are shown in the Appendix (Figure~\ref{fig:extra_cmd}). These CMDs distinguish between the main-sequence and giant branches and thus clean the sample from most evolved stars. Main-sequence candidates are selected as lying within the region where the CMD density prior is above $0.08$, as are any stars with $M_r > 7$\,AB\,mag.  Stars that fall within this boundary {\em in all three CMDs} are then classified as high-confidence dC stars.  Figure~\ref{fig:CMD} is shown as an example, where blue points within the border are stars excluded owing to the constraints in the other two colours. This results in a sample of 1003 high-confidence dC stars.

\begin{figure*}
    \centering
    \includegraphics[width=\linewidth]{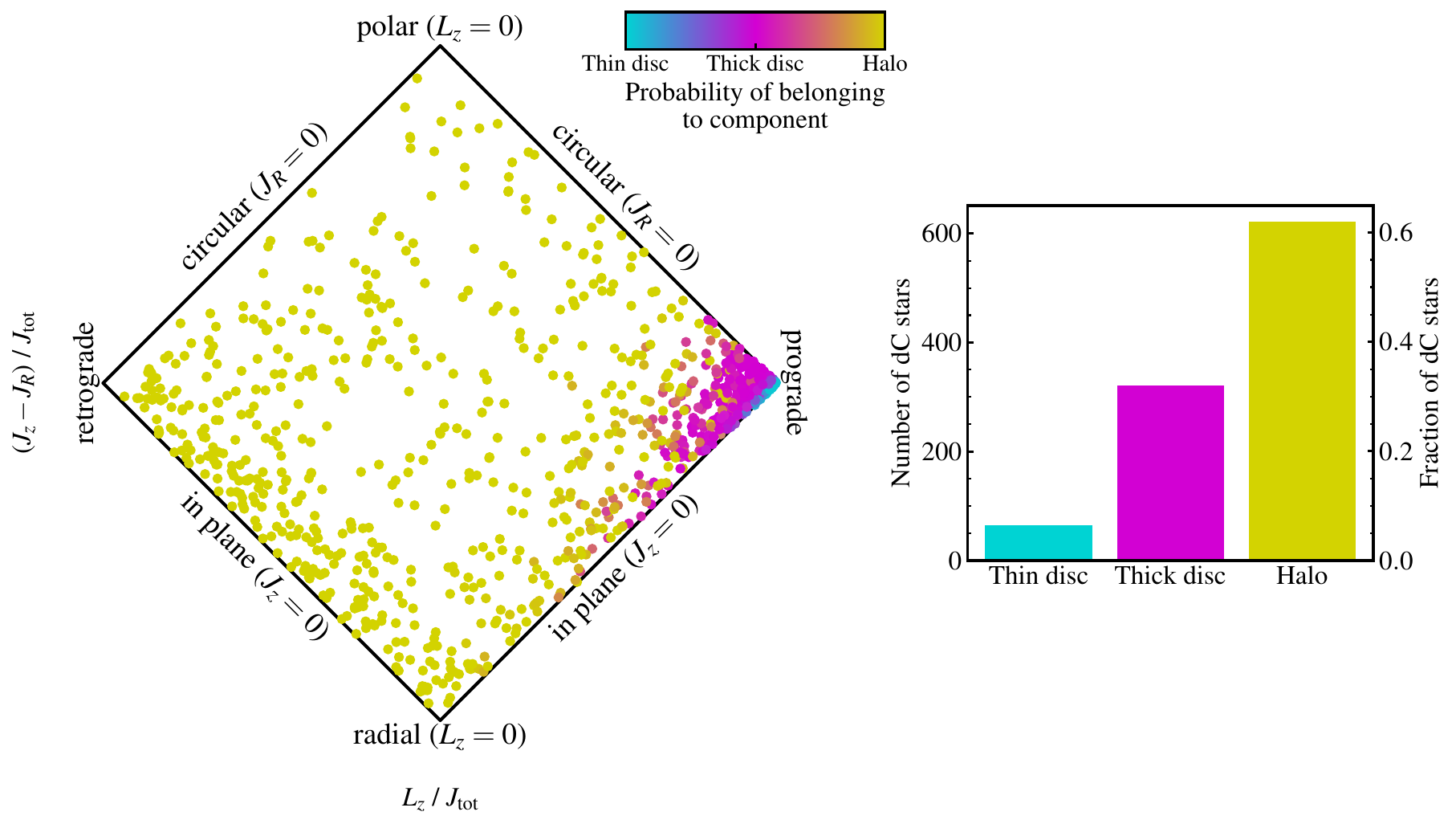}
    \caption{\textit{Left:} The action-space map of the 1003 high-confidence dC stars. The horizontal axis represents the normalized angular momentum in the (vertical) Galactic $Z$-direction, ranging from $-1$ for retrograde to $+1$ for prograde orbits.  The vertical axis represents the relation between the vertical and the radial action. Objects on the lower edges move in the Galactic plane, while stars located toward the upper corner move on increasingly polar orbits. Stars located along the upper edges have circular orbits, but these become increasingly radial for diagram positions toward the bottom corner. Therefore, stars located in the right corner move within the disc (prograde, in the plane, and circular), while stars positioned away from this position have thick disc and halo orbits.  The dC stars are colour-coded by their median probabilities of belonging to each Galactic component, with a continuous distribution of colours, representing stars on the edge between two components; stars in violet are between the thin and thick disc, and stars in orange are between the thick disc and halo. \textit{Right:} Distribution of dC stars in the different Galactic components resulting from the action-based classification: the thin disc contains 6.5\,per cent (59 stars), the thick disc 31.4\,per cent (320 stars), and the halo 62.1\,per cent (624 stars) of the dC population.  This is a strong indication of an old and metal-poor population.}
    \label{fig:actions}
\end{figure*}

The distribution of dC stars in absolute magnitude reveals that the population likely possesses a relatively broad continuum of masses and effective temperatures.  Furthermore, despite the prior pulling stars toward the stellar locus, a significant fraction of the population lies to the left of the main sequence. This is illustrated in the bottom panel of Figure~\ref{fig:CMD}, which shows the difference in the absolute magnitude with respect to the mode of the prior at each $(r-i)$. Clearly, the bulk of the dC candidates are fainter than the typical main-sequence star at fixed colour. Comparison with a set of model stellar isochrones \citep{Dotter2008} demonstrates that this behaviour is expected for populations of $\mathrm{[Fe/H]}\sim-1\,\mathrm{dex}$. Although the isochrones are for carbon-normal stars, the $r$ and $i$ bands are expected to be less affected by the carbon-rich nature than the $g$ band, which overlaps with the series of C$_2$ Swan bands. This comparison indicates that the population as a whole is likely to be predominantly metal-poor. As previously mentioned, the only dC star with a constrained metallicity is G77-61 ([Fe/H]\,$=-4.0$; \citealt{Gass1988,Plez2005}), and remains one of the most metal-deficient stars known.

A final consideration is whether the adoption of the photometric prior has biased the dC distances. As already noted, it appears the typical dC star is fainter than the typical main-sequence star, so in the absence of strong information from the parallaxes, the distances from this method will be biased towards larger values. The lower panel of Figure~\ref{fig:CMD} shows the main sequence offset for the full sample coloured by parallax signal-to-noise, $\varpi/\upsigma_\varpi$, along with median trends for the full sample and those with $\varpi/\upsigma_\varpi>20$. There is a clear trend with $\varpi/\upsigma_\varpi$, where more accurate parallaxes fall (on average) further below the main sequence (and indeed agree well with the $\mathrm{[Fe/H]}=-1\,\mathrm{dex}$ isochrone), while those with lower $\varpi/\upsigma_\varpi$ lie closer to the main-sequence prior. At most, this effect appears to be $\updelta M_{r,0}\approx1.5$\,AB\,mag for stars with $(r-i)\approx0.7$, which would translate into distance errors of approximately 70\,per cent. However, for stars around $(r-i)\approx0.4$, the bias is of the same order as the random error.  This bias will incorrectly place the stars further from the sun and increase the inferred tangential velocity.

\section{Dynamical classification of \texorpdfstring{\lowercase{d}C}{dC} stars}

Without information on metallicity or abundance for the dC stars, Galactic dynamics is used to constrain the fraction of the population that is likely to be old and metal-poor. The bulk of stars in each major component of the Galaxy have distinct, characteristic ages and chemical properties: broadly, the thin disc is the youngest and most metal-rich, while stars in the halo are, on average, older and metal-poor, with stars in the thick disc somewhere in between.  The distribution of the dC stars over these Galactic components should then trace their bulk metallicity and age properties.

\subsection{Component assignment using distribution functions}\label{sec:class:method}

Full knowledge of the 6-dimensional position and velocity, combined with an underlying gravitational potential of the Milky Way, yields the Galactic orbit of any star, which can be labelled by actions. Distribution functions describe the probability that a star belonging to a given component is found on the orbit described by its actions.  The three components of interest, the thin disc, thick disc, and stellar halo, are described with published model distributions as follows. 

\citet{Binney2023} present a self-consistent Galaxy model, where the gravitational potential is computed from the sub-component distribution functions. The Galactic components fitted in that work are three thin discs (where age is used in labelling but not in the modelling), a thick disc, a spheroidal bulge, and spheroidal stellar and dark matter halos.  The stellar discs are described as exponential distributions, and the spheroidal components as double power-laws. The distribution functions of interest are the thin disc (summed over the three model components), the thick disc, and the stellar halo. These distribution functions have been fitted to {\em Gaia} DR2 data using {\sc agama} \citep{Vasiliev2019}.  One caveat is that for technical reasons related to the density of stars at $J_\phi=0$, the halo distribution function employed by \citet{Binney2023} has no significant radial bias contrary to observations. This may weakly bias dC stars on more circular orbits to be assigned to the halo component rather than the thick disc.

To include the observational uncertainties, for each star, 1000 random realizations of $(\upalpha,\updelta, \mu_\upalpha,\mu_\updelta)$ are drawn from their correlated uncertainty distributions, while radial velocities are sampled from the SEGUE data described in Section~\ref{sec:data}, and distances from their posterior chains (\ref{sec:dist:res}).  The radial velocities of all 1003 high-confidence dC stars are found to be reliable based on the SEGUE quality flag \texttt{zwarning\_elodie}.  Note that, although the true proper motions are inferred in Section~\ref{sec:distances}, these are not used here.  For each realization of each star, the actions ($J_r$, $J_z$, $J_\phi=L_z$) and the probability of belonging to each Galactic component are calculated. Unbound posterior samples have undefined actions so their Galactic component membership under an equilibrium distribution function is ambiguous: these posterior samples are defined as belonging to the halo component. The component probabilities are individually normalized by the sum of all three components, where the median, $16^{\rm th}$, and $84^{\rm th}$ percentiles are used to obtain probabilities and actions for each star, with uncertainties. Each dC star is then assigned to the Galactic component with the highest probability.

\subsection{Fraction of halo, thick and thin disc orbits}

\begin{figure}
    \centering
    \includegraphics[width=\columnwidth]{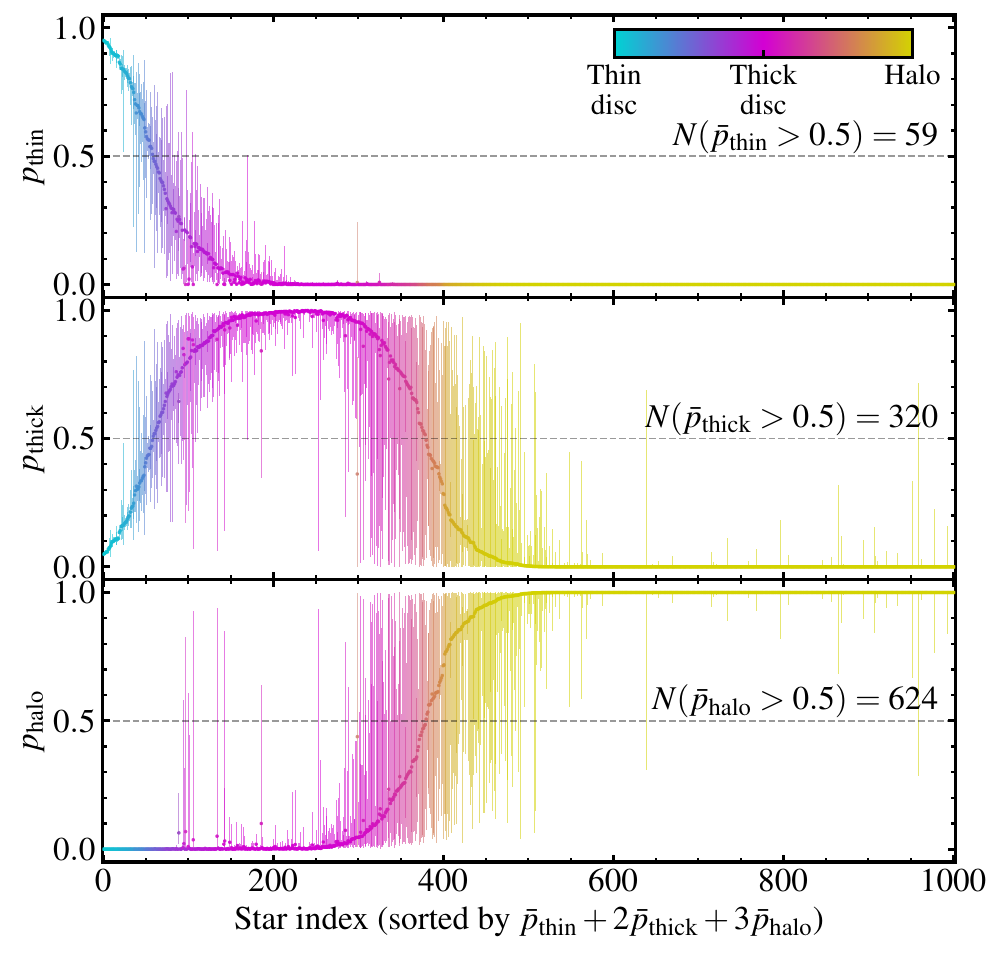}
    \caption{Individual Galactic component posteriors for the 1003 high-confidence dC stars. Each panel shows a dot and bar per star representing the posterior median and the 16$^{\rm th}$ to 84$^{\rm th}$ percentile confidence range, respectively, for membership in each component (top: thin disc; middle: thick disc; bottom: halo). The stars are ordered horizontally by the sum of median probabilities, $\bar{p}_\mathrm{thin}+2\bar{p}_\mathrm{thick}+3\bar{p}_\mathrm{halo}$, and coloured by their median classification as per the colourbar. The annotation in each panel gives the number of stars with a median classification probability greater than $0.5$ for each component. It can be seen that a large fraction of stars classified as halo members have an almost 100\,per cent probability of belonging to the halo, and almost zero probability of belonging to the two disc components. In contrast, the distribution of stars classified as thin disc is not as clean, where many objects have a non-negligible probability of belonging to the thick disc. Distinguishing between the thin and thick disc solely based on dynamics is not straightforward, where chemical abundances would improve upon this purely dynamical analysis. However, the classification is reliable for thick disc and particularly halo stars.
    }
    \label{fig:components_prob}
\end{figure}

With the stars assigned to the different Galactic components, their properties are now explored. The right-hand panel of Figure~\ref{fig:actions} shows the number of dC stars with the highest assigned probability in each component, where ($62.1\pm 0.6$)\,per cent ($N=624$) of the population belong to the stellar halo, ($31.4\pm0.7$)\,per cent ($N=320$) are part of the thick disc, and only ($6.5\pm0.4$)\,per cent ($N=59$) are members of the thin disc (where uncertainties are derived from standard deviations over posterior samples for each star). The median Galactic component posterior probabilities, $\bar{p}_i$, and their corresponding $16^{\rm th}$ to $84^{\rm th}$ confidence intervals are shown in Figure~\ref{fig:components_prob}. Here the stars are ordered horizontally by $\bar{p}_\mathrm{thin}+2\bar{p}_\mathrm{thick}+3\bar{p}_\mathrm{halo}$ to sort confident thin disc members on the left and confident halo members on the right. It can be seen that a significant fraction of the sample is dynamically classified with high confidence. There is some ambiguity between the thick disc and halo for $\sim10$\,per cent of the sample, arising from the propagation of observational uncertainties. On the left-hand side of the diagram, there are stars with tight posteriors owing to small observational uncertainties, for which there is some ambiguity between thin and thick disc membership because of their overlapping dynamical distribution functions.

The distribution functions depend on the actions of each star, which are visualised in Figure~\ref{fig:actions}. This action-space map of the dC stars, where each object is colour-coded by the probability of belonging to each component, shows how the actions inform the distribution functions. The distribution of thin disc stars is tightly confined to the right corner, with thick disc stars spreading further along the in-plane axis, and toward more radial orbits. The remaining objects in the diagram, and thus the bulk of dC stars, are dominated by those with halo orbits. The analysis in this work has taken care to propagate the uncertainties in the actions (and thus the classification) of each star. In Appendix~\ref{app:plots}, the distribution of the actions along with their uncertainties are shown. Although these uncertainties can be sizeable, it is clear they are sufficiently small for an accurate classification of each star.

\section{Discussion}

The results here provide the first substantial catalogue of high-confidence dC stars, where the dynamics suggest they are predominantly a halo population, likely implying that they are also a metal-poor or an old population. These results are discussed in relation to previous work, with some caveats and possible implications.

\subsection{Comparison with previous work} 

Based on the preceding analysis, only a small percentage of dC stars are likely thin disc members, while the remaining fraction have orbits consistent with a metal-poor population, dominated by objects that were formed in the Galactic halo. These results are consistent with a prior study using the same SDSS sample of dC stars as in this work, based on proper motion catalogues generated from {\em Gaia} DR1 astrometry, finding between 30 and 60\,per cent of dC candidate stars belong to the stellar halo \citep{Farihi2018}. That pre-\textit{Gaia} DR2 analysis relies on both reduced proper motion with isochrone fitting, as well as Toomre diagrams, to identify the different Galactic components kinematically \citep[see e.g.][]{Venn2004}, assuming that all dC stars have $M_r = 8.0$\,AB\,mag. This bulk absolute magnitude estimate agrees remarkably well with the results of the sample in this paper, which has a mean (median) absolute magnitude $M_r = 7.7$\,AB\,mag (but a substantial spread, Figure~\ref{fig:CMD}).

\begin{figure}
    \centering
    \includegraphics[width=\columnwidth]{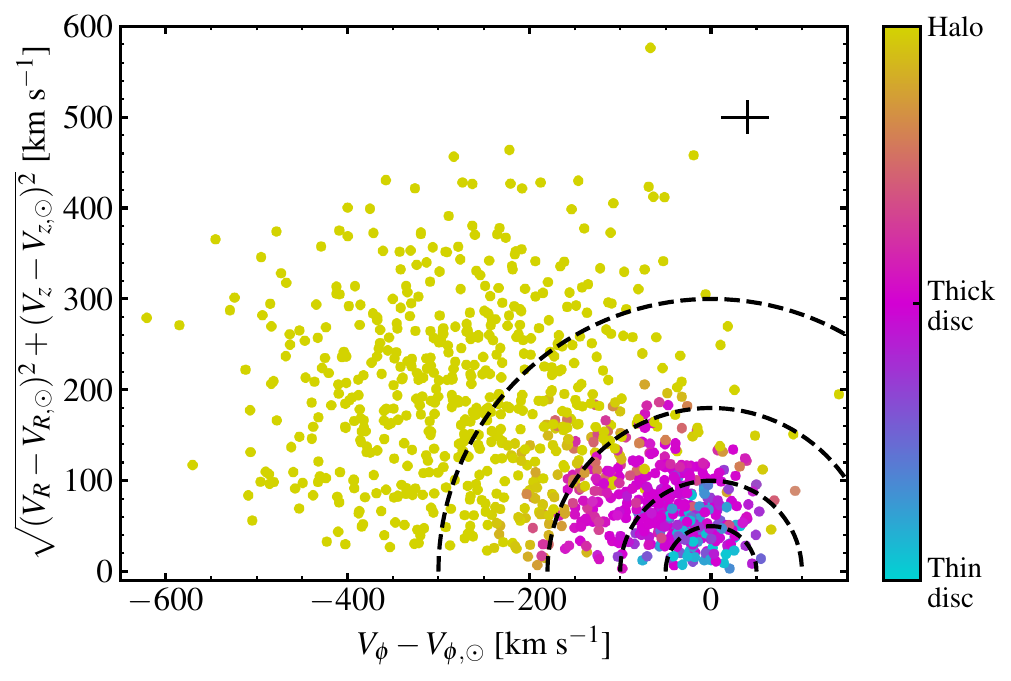}
    \caption{Toomre diagram of the high-confidence dC stars colour-coded as in Figure~\ref{fig:actions}.  The diagram is corrected to the local standard of rest, so that the thin disc has $V_\phi-V_{\phi,\odot}=0$\kms.  The dashed semi-circles trace constant total space velocities of 50, 100, 180, and 300\kms. The error bar in the upper right shows the median uncertainty. While the distribution of dC stars in these Galactic components roughly follows the semi-circular benchmarks, owing to the broad spatial distribution, several stars are in locations where they would be attributed to a different component if the selection is done only kinematically using the Toomre diagram.}
    \label{fig:toomre}
\end{figure}

More recently and post-{\em Gaia} DR2, \citet{Roulston2022} use a Toomre diagram for estimating the age for a subsample of the SDSS dC stars that are also analyzed here.  In that study, the dC stars are compared to a sample of carbon-normal K and M dwarfs that are reported to represent the thin and thick disc, and the authors subsequently argue that the resulting poor match implies the dC stars have either a thick disc or halo origin. To underscore the success of the dynamical analysis shown in Figure~\ref{fig:actions} as compared to a purely kinematic selection, the 1003 dC stars are plotted in a Toomre diagram in Figure~\ref{fig:toomre}, colour-coded by their component probabilities.  The components roughly segregate along the curves of constant velocity, with some overlap between.  Notably, however, the sample distributions in Figure~\ref{fig:toomre} and those in \citet{Roulston2022} do not match; e.g.\ their figure~1 does not display any stars in the retrograde disc (i.e.\ at low $\sqrt{U^2 + W^2}$ and negative $V$). This could be due to an incorrect coordinate transformation of their data.

It has been suggested that the kinematics of dC stars might be biased owing to radial velocities that include a significant orbital motion component for those in close binaries, such as the handful of systems consistent with a post-common envelope configuration \citep{Margon2018, Whitehouse2021, Roulston2022}.  However, existing radial velocity measurements are consistent with circular orbits, where the average orbital speed along the line of sight is zero, and there is an equal chance to decrease or increase the total observed line-of-sight velocity.  Furthermore, the orbital velocity semi-amplitudes for dC stars in these binaries are typically smaller than 50\kms, with a few exceptions closer to 100\kms, and a random measurement will yield no more than 0.72 of these extrema.  The dominant halo population of dC stars have space velocities that are significantly larger than the average range of orbital speeds measured along the line of sight, and therefore, based on this and the aforementioned caveats, orbital motion cannot significantly inflate the inferred kinematics of dC stars.

The results found here for dC stars passing through the solar neighborhood are directly comparable to a kinematical study of 644 CEMP giant stars situated closer to and within the Galactic halo \citep[][their figure 8 is directly comparable to Figures~\ref{fig:actions} and \ref{fig:ELz}]{Zepeda2023}.  While that study leverages chemical abundance data that are not available for the dC sample, they find the CEMP stars are distributed over the entirety of orbital action space, similar to the dC population with the exception of its disk component.  Abundance information allows co-identification with halo substructures for a few dozen CEMP stars, which is not possible here, but nevertheless the dC halo stars overlap with known substructures in a similar manner.  Future data and modeling are necessary for definitive substructure association for any dC stars.

\subsection{Selection function of the sample}

To assess the impact of this spatial selection, the component membership probabilities are computed from the \citet{Binney2023} model using only the spatial locations of the sample (i.e.\ the density fractions of each component). For the entire sample, the ratio of thin:thick:halo membership based purely on spatial location is found to be $0.17:0.67:0.16$, while additionally using the kinematics gives $0.07:0.31:0.62$. There is indeed a bias toward observing thick disc and halo objects given the SEGUE selection but despite this, the dynamic classification shows the sample has a stronger tendency toward the thick disc and halo than expected purely on the basis of the selection. This is evidence that the bulk of dC stars indeed have thick disc or halo kinematics, and are thus metal-poor or old, but fuller modelling of the selection function would be required to conclude this definitively.

\subsection{Do thick disc and halo kinematics imply metal-poor or old?}

The dC sample has been shown to have kinematics dominated by stars consistent with membership in the thick disc or halo.  This was assessed using purely dynamical distribution functions designed to give the best representation of the density and kinematics of the Galaxy with no explicit consideration of the properties of the stars \citep{Binney2023}. However, it is well known that the average star in these Galactic components \emph{is metal-poor and old}, with ${\rm [Fe/H]}\lesssim-0.5$ (consistent with the isochrone comparison in Figure~\ref{fig:CMD}), and age $\gtrsim10$\,Gyr \citep[e.g.][]{Ivezic2008, Kilic2017}, thus the association between metallicity-age and kinematics.  

\begin{figure}
    \centering
    \includegraphics[width=\columnwidth]{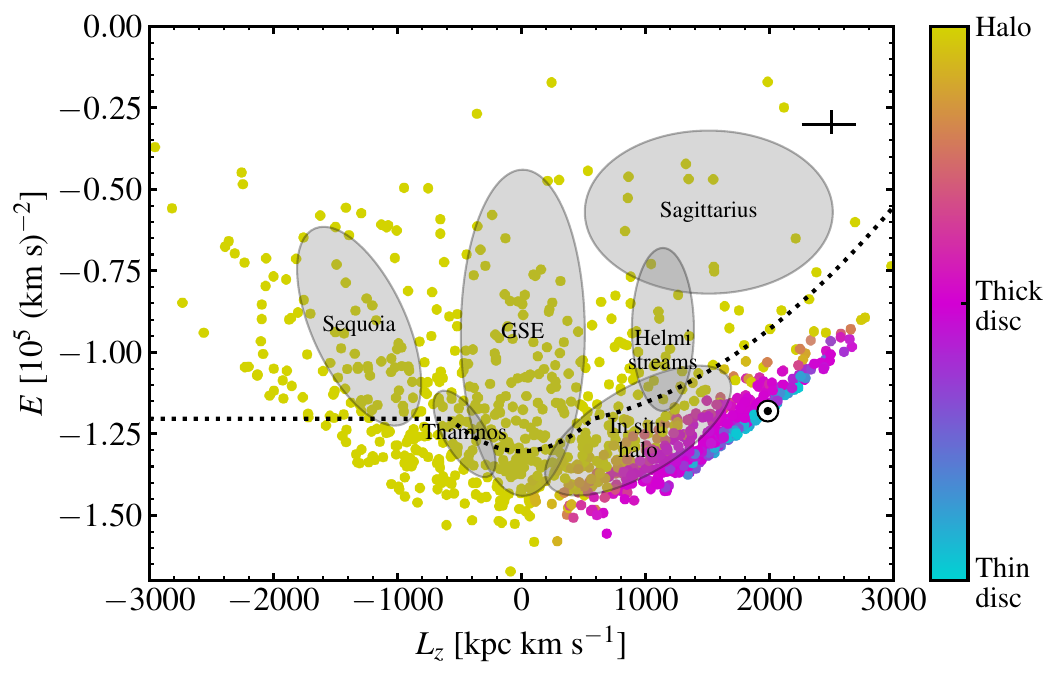}
    \caption{Energy-angular momentum diagram of the 1003 high-confidence dC stars colour-coded as in Figure~\ref{fig:actions}. The conventional sun symbol plots the location of a circular orbit at the solar Galacto-centric radius. The error bar in the upper right is the median uncertainty of the sample. The dashed black line marks the approximate separation between in-situ and accreted globular clusters \citep{Belokurov2024}, and the grey labelled ellipses show sub-structures \citep{Naidu2020}.}
    \label{fig:ELz}
\end{figure}

Recent work and the arrival of {\it Gaia} data have enabled a richer and deeper understanding of the Milky Way stellar halo.  Figure~\ref{fig:ELz} plots the energy-angular momentum diagram for the high-confidence dC stars along with the \emph{in-situ}-accreted boundary for globular clusters proposed by \citet{Belokurov2024}, and the approximate locations of the structures classified by \citet{Naidu2020}.  It can be seen that there is no particularly strong tendency toward any particular halo sub-component in the sample and the stars classified as halo are distributed across both accreted and \emph{in-situ} parts of the diagram. 

The association of dC stars with different halo sub-components would have implications for their metallicities and ages.  For example, the GSE stars are known to have $-1.28 < {\rm [Fe/H]} < -1.18$ with a low metallicity tail down to at least ${\rm [Fe/H]} = -2.1$ \citep{Amarante2022, Carrillo2024}, and are predominantly old with a possible tail of younger stars \citep{Horta2024}. In contrast, the in-situ halo component \citep{Bonaca2017, Belokurov2020} can contain old stars with $\mathrm{[Fe/H]}\approx-0.7$, and highly radial and possibly even retrograde orbits. Further study is warranted to more accurately classify the sub-structure membership of the sample studied here, but based on the energy-angular momentum diagram and comparison with known sub-structures, the dC stars seem fairly representative of the average local halo star. It is thus likely that, on average, the sample is old and metal-poor.

A further tool for assessing more explicitly the metallicity and ages of the stars is using an \emph{extended} distribution function \citep[e.g.][]{SandersBinney2015, BinneyVasiliev2025}. These models attempt to reproduce both the dynamical \emph{and} chemical structure of the Galaxy, so one can construct e.g. $p([\mathrm{Fe/H}] |\, \bm{x},\bm{v})$. Alternatively, data-driven label transfer techniques that attempt to match kinematics as a function of auxiliary parameters could be employed \citep{Zhang2025}. These are both potentially fruitful directions for further work with this sample.

\subsection{The origin of dC halo stars}\label{sec:disc:origin}

Metal-deficient stars are more susceptible to atmospheric pollution via mass transfer (e.g.\ metal-free white dwarfs), because only a fraction of carbon-rich material is sufficient to increase the C/O ratio beyond 1, relative to solar metallicity counterparts \citep{deKool1995}, all else being equal.  This is part of the consensus interpretation for CH and CEMP-s stars, which are low-metallicity binaries where the secondary star is carbon-enhanced post-mass transfer \citep[e.g.][]{Lucatello2005, Jorissen2016}.  The likely metal-poor nature of the dC population implied by the dynamical analysis presented in this work is consistent with a similar, binary origin, at least for some fraction, and where a growing body of work suggests dC stars are often found in binaries where past mass transfer is likely or plausible \citep{Whitehouse2018,Harris2018,Margon2018,Roulston2021}.  

It remains unknown if there are dC stars analogous to the CEMP-no population, which may contain intrinsically carbon-rich stars that are not correlated with their binary properties (similar to the nearby field population; \citealt{Starkenburg2014, Yoon2016}).  Only G77-61 has detailed abundances for such a comparison, and it is often classified as CEMP-no based on the non-detection of barium, and its extreme metal poverty, but it should be noted the upper limit on s- and r-process elements is not as robust as for typical (giant) CEMP stars \citep{Plez2005,Arentsen2019}.  Based on Figure~\ref{fig:CMD}, the dC stars in this study span a mass range that likely overlaps with the known CEMP population (which are giants owing to luminosity bias), but also extends to lower masses that will not leave the main sequence in a Hubble time.

Nevertheless, the results have also demonstrated that there is a subdominant thin disc component in the dC sample. Thin disc stars are characteristically solar metallicity so it is significantly more difficult to produce a carbon-rich atmosphere through mass transfer for these objects. However, the results suggest it is possible in rare cases (especially considering the high ratio of solar metallicity to metal-poor star populations). Further study of dC stars in more metal-rich environments and a more detailed characterization of the selection function can help constrain their binary properties and the efficiency of mass transfer as a function of metallicity.

\section{Conclusions and Outlook}

This study has analyzed 1003 high-confidence dC stars to determine their distances, 3-dimensional space motions, and Galactic orbits, using a combination of SDSS DR16 photometry and radial velocity data, and state-of-the-art \textit{Gaia} DR3 astrometry.  These properties are used to dynamically classify the dC stars and assign component membership likelihood, using action-based distribution functions.  This is achieved by first using Bayesian inference to obtain accurate distances for these faint stars, using a set of astrometric and photometric priors based on expected HR diagram positions. The results are summarized as follows:

\begin{itemize}

    \item Probabilistic distances are inferred for 1208 candidate dC stars from \citet{Green2013}, using astrometry from \textit{Gaia} DR3 and photometry from SDSS DR16 (Figure~\ref{fig:distances}). With these distances, colour-absolute magnitude diagrams are computed for all candidate stars. By selecting regions in the $(g-r)$, $(g-i)$, and $(r-i)$ colours that cover the main sequence and below, a total of 1003 high-confidence dC stars are identified (Figure~\ref{fig:CMD}).

    \item The actions of the 1003 dC stars with reliable astrometry and spectroscopy are calculated, and Galactic component membership is assigned based on the distribution functions from \citet{Binney2023}.  Each star is thus assigned to the thin disc, the thick disc, or the stellar halo.  Roughly 60\,per cent of the dC population are found to have dynamics consistent with halo membership and around 30\,per cent with thick disc orbits (Figures~\ref{fig:actions} and \ref{fig:components_prob}). It is demonstrated that the conclusion of predominantly halo membership is not a result of the spatial selection function of the sample.

    \item The high-confidence dC stars preferentially fall below (to the left of) the main sequence in the $M_r$ vs.\ $(r-i)$ diagram. Comparison with carbon-normal isochrones in this plane suggests the dC population is metal poor with $\mathrm{[Fe/H]}\sim-1$. The dynamical association with the thick disc and halo yields further evidence that the dC stars are metal-poor or old. Comparison with known Galactic halo sub-structures does not reveal a tendency toward any known sub-component, suggesting they are broadly representative of the average metal-poor and old stellar halo (Section~\ref{sec:disc:origin}).

\end{itemize}

These results demonstrate the significance of further investigation of these rare, poorly understood stars. The next generation of large spectroscopic surveys (e.g.\ 4MOST) will likely observe many known dC stars and possibly discover new examples. Furthermore, the results suggest that a more dedicated spectroscopic follow-up campaign of dC stars should be undertaken.  Spectroscopic analysis to measure metallicities and abundances of these stars remains a challenge, but the rewards of overcoming current limitations are tempting: a detailed spectroscopic study would provide conclusive evidence on the metallicity of the population as a whole, possibly opening up new avenues to discover and study very (or extremely) metal-poor stars.

\section*{Acknowledgements}
The authors thank an anonymous reviewer for their feedback, Eugene Vasiliev for help with the {\sc agama} software, and Adam Dillamore for help annotating the energy-angular momentum diagram. This project was developed in part at the {\it Gaia} F\^ete, hosted by the Flatiron Institute Center for Computational Astrophysics in 2022 June.  JLS acknowledges the support of the Royal Society (URF\textbackslash R1\textbackslash191555; URF\textbackslash R\textbackslash 241030).  This work has made use of data from the European Space Agency {\it Gaia} mission, and the Data Processing and Analysis Consortium, with funding provided by national institutions.

\section*{Data Availability}
The data used in this paper are available in public archives from the Sloan Digital Sky Survey and the European Space Agency. The dataset produced in this study is made available alongside this publication with the data columns described in Table~\ref{tab:data}. The table and the code to reproduce all calculations and plots in the paper are provided at \url{https://github.com/jls713/dc}.




\bibliographystyle{mnras}
\bibliography{references} 

\begin{thebibliography}{}
\makeatletter
\relax
\def\mn@urlcharsother{\let\do\@makeother \do\$\do\&\do\#\do\^\do\_\do\%\do\~}
\def\mn@doi{\begingroup\mn@urlcharsother \@ifnextchar [ {\mn@doi@}
  {\mn@doi@[]}}
\def\mn@doi@[#1]#2{\def\@tempa{#1}\ifx\@tempa\@empty \href
  {http://dx.doi.org/#2} {doi:#2}\else \href {http://dx.doi.org/#2} {#1}\fi
  \endgroup}
\def\mn@eprint#1#2{\mn@eprint@#1:#2::\@nil}
\def\mn@eprint@arXiv#1{\href {http://arxiv.org/abs/#1} {{\tt arXiv:#1}}}
\def\mn@eprint@dblp#1{\href {http://dblp.uni-trier.de/rec/bibtex/#1.xml}
  {dblp:#1}}
\def\mn@eprint@#1:#2:#3:#4\@nil{\def\@tempa {#1}\def\@tempb {#2}\def\@tempc
  {#3}\ifx \@tempc \@empty \let \@tempc \@tempb \let \@tempb \@tempa \fi \ifx
  \@tempb \@empty \def\@tempb {arXiv}\fi \@ifundefined
  {mn@eprint@\@tempb}{\@tempb:\@tempc}{\expandafter \expandafter \csname
  mn@eprint@\@tempb\endcsname \expandafter{\@tempc}}}

\bibitem[\protect\citeauthoryear{{Abate}, {Pols}, {Stancliffe}, {Izzard},
  {Karakas}, {Beers}  \& {Lee}}{{Abate} et~al.}{2015}]{Abate2015}
{Abate} C.,  {Pols} O.~R.,  {Stancliffe} R.~J.,  {Izzard} R.~G.,  {Karakas}
  A.~I.,  {Beers} T.~C.,   {Lee} Y.~S.,  2015, \mn@doi [\aap]
  {10.1051/0004-6361/201526200}, \href
  {https://ui.adsabs.harvard.edu/abs/2015A&A...581A..62A} {581, A62}

\bibitem[\protect\citeauthoryear{{Ahumada} et~al.,}{{Ahumada}
  et~al.}{2020}]{SDSS_DR16}
{Ahumada} R.,  et~al., 2020, \mn@doi [\apjs] {10.3847/1538-4365/ab929e}, \href
  {https://ui.adsabs.harvard.edu/abs/2020ApJS..249....3A} {249, 3}

\bibitem[\protect\citeauthoryear{{Amarante}, {Debattista}, {Beraldo e Silva},
  {Laporte}  \& {Deg}}{{Amarante} et~al.}{2022}]{Amarante2022}
{Amarante} J. A.~S.,  {Debattista} V.~P.,  {Beraldo e Silva} L.,  {Laporte} C.
  F.~P.,   {Deg} N.,  2022, \mn@doi [\apj] {10.3847/1538-4357/ac8b0d}, \href
  {https://ui.adsabs.harvard.edu/abs/2022ApJ...937...12A} {937, 12}

\bibitem[\protect\citeauthoryear{{Arentsen}, {Starkenburg}, {Shetrone}, {Venn},
  {Depagne}  \& {McConnachie}}{{Arentsen} et~al.}{2019}]{Arentsen2019}
{Arentsen} A.,  {Starkenburg} E.,  {Shetrone} M.~D.,  {Venn} K.~A.,  {Depagne}
  {\'E}.,   {McConnachie} A.~W.,  2019, \mn@doi [\aap]
  {10.1051/0004-6361/201834146}, \href
  {https://ui.adsabs.harvard.edu/abs/2019A&A...621A.108A} {621, A108}

\bibitem[\protect\citeauthoryear{{Arentsen}, {Placco}, {Lee}, {Aguado},
  {Martin}, {Starkenburg}  \& {Yoon}}{{Arentsen} et~al.}{2022}]{Arentsen2022}
{Arentsen} A.,  {Placco} V.~M.,  {Lee} Y.~S.,  {Aguado} D.~S.,  {Martin} N.~F.,
   {Starkenburg} E.,   {Yoon} J.,  2022, \mn@doi [\mnras]
  {10.1093/mnras/stac2062}, \href
  {https://ui.adsabs.harvard.edu/abs/2022MNRAS.515.4082A} {515, 4082}

\bibitem[\protect\citeauthoryear{{Aringer}, {Girardi}, {Nowotny}, {Marigo}  \&
  {Bressan}}{{Aringer} et~al.}{2016}]{Aringer2016}
{Aringer} B.,  {Girardi} L.,  {Nowotny} W.,  {Marigo} P.,   {Bressan} A.,
  2016, \mn@doi [\mnras] {10.1093/mnras/stw222}, \href
  {https://ui.adsabs.harvard.edu/abs/2016MNRAS.457.3611A} {457, 3611}

\bibitem[\protect\citeauthoryear{{Astraatmadja} \&
  {Bailer-Jones}}{{Astraatmadja} \& {Bailer-Jones}}{2016}]{Astraatmadja2016}
{Astraatmadja} T.~L.,  {Bailer-Jones} C. A.~L.,  2016, \mn@doi [\apj]
  {10.3847/0004-637X/832/2/137}, \href
  {https://ui.adsabs.harvard.edu/abs/2016ApJ...832..137A} {832, 137}

\bibitem[\protect\citeauthoryear{{Bailer-Jones}}{{Bailer-Jones}}{2011}]{Bailer-Jones2011}
{Bailer-Jones} C.~A.~L.,  2011, \mn@doi [\mnras]
  {10.1111/j.1365-2966.2010.17699.x}, \href
  {https://ui.adsabs.harvard.edu/abs/2011MNRAS.411..435B} {411, 435}

\bibitem[\protect\citeauthoryear{{Bailer-Jones}}{{Bailer-Jones}}{2015}]{Bailer-Jones2015}
{Bailer-Jones} C. A.~L.,  2015, \mn@doi [\pasp] {10.1086/683116}, \href
  {https://ui.adsabs.harvard.edu/abs/2015PASP..127..994B} {127, 994}

\bibitem[\protect\citeauthoryear{Bailer-Jones}{Bailer-Jones}{2017}]{Bailer-Jones2018}
Bailer-Jones C.,  2017, {J}oint inference from parallax and proper motions,
  Provided by ESA, \url
  {https://dms.cosmos.esa.int/COSMOS/doc_fetch.php?id=3504158}

\bibitem[\protect\citeauthoryear{{Bailer-Jones}, {Rybizki}, {Fouesneau},
  {Demleitner}  \& {Andrae}}{{Bailer-Jones} et~al.}{2021}]{Bailer-Jones2021}
{Bailer-Jones} C.~A.~L.,  {Rybizki} J.,  {Fouesneau} M.,  {Demleitner} M.,
  {Andrae} R.,  2021, \mn@doi [\aj] {10.3847/1538-3881/abd806}, \href
  {https://ui.adsabs.harvard.edu/abs/2021AJ....161..147B} {161, 147}

\bibitem[\protect\citeauthoryear{{Beers} \& {Christlieb}}{{Beers} \&
  {Christlieb}}{2005}]{Beers2005}
{Beers} T.~C.,  {Christlieb} N.,  2005, \mn@doi [\araa]
  {10.1146/annurev.astro.42.053102.134057}, \href
  {https://ui.adsabs.harvard.edu/abs/2005ARA&A..43..531B} {43, 531}

\bibitem[\protect\citeauthoryear{{Belokurov} \& {Kravtsov}}{{Belokurov} \&
  {Kravtsov}}{2024}]{Belokurov2024}
{Belokurov} V.,  {Kravtsov} A.,  2024, \mn@doi [\mnras]
  {10.1093/mnras/stad3920}, \href
  {https://ui.adsabs.harvard.edu/abs/2024MNRAS.528.3198B} {528, 3198}

\bibitem[\protect\citeauthoryear{{Belokurov} et~al.,}{{Belokurov}
  et~al.}{2020}]{Belokurov2020}
{Belokurov} V.,  et~al., 2020, \mn@doi [\mnras] {10.1093/mnras/staa1522}, \href
  {https://ui.adsabs.harvard.edu/abs/2020MNRAS.496.1922B} {496, 1922}

\bibitem[\protect\citeauthoryear{{Binney} \& {Vasiliev}}{{Binney} \&
  {Vasiliev}}{2023}]{Binney2023}
{Binney} J.,  {Vasiliev} E.,  2023, \mn@doi [\mnras] {10.1093/mnras/stad094},
  \href {https://ui.adsabs.harvard.edu/abs/2023MNRAS.520.1832B} {520, 1832}

\bibitem[\protect\citeauthoryear{{Binney} \& {Vasiliev}}{{Binney} \&
  {Vasiliev}}{2024}]{BinneyVasiliev2025}
{Binney} J.,  {Vasiliev} E.,  2024, \mn@doi [\mnras] {10.1093/mnras/stad3312},
  \href {https://ui.adsabs.harvard.edu/abs/2024MNRAS.527.1915B} {527, 1915}

\bibitem[\protect\citeauthoryear{{Bonaca}, {Conroy}, {Wetzel}, {Hopkins}  \&
  {Kere{\v{s}}}}{{Bonaca} et~al.}{2017}]{Bonaca2017}
{Bonaca} A.,  {Conroy} C.,  {Wetzel} A.,  {Hopkins} P.~F.,   {Kere{\v{s}}} D.,
  2017, \mn@doi [\apj] {10.3847/1538-4357/aa7d0c}, \href
  {https://ui.adsabs.harvard.edu/abs/2017ApJ...845..101B} {845, 101}

\bibitem[\protect\citeauthoryear{{Carrillo}, {Deason}, {Fattahi}, {Callingham}
  \& {Grand}}{{Carrillo} et~al.}{2024}]{Carrillo2024}
{Carrillo} A.,  {Deason} A.~J.,  {Fattahi} A.,  {Callingham} T.~M.,   {Grand}
  R. J.~J.,  2024, \mn@doi [\mnras] {10.1093/mnras/stad3274}, \href
  {https://ui.adsabs.harvard.edu/abs/2024MNRAS.527.2165C} {527, 2165}

\bibitem[\protect\citeauthoryear{{Christlieb}, {Wisotzki}  \&
  {Gra{\ss}hoff}}{{Christlieb} et~al.}{2002}]{Christlieb2002}
{Christlieb} N.,  {Wisotzki} L.,   {Gra{\ss}hoff} G.,  2002, \mn@doi [\aap]
  {10.1051/0004-6361:20020830}, \href
  {https://ui.adsabs.harvard.edu/abs/2002A&A...391..397C} {391, 397}

\bibitem[\protect\citeauthoryear{{Cooke}, {Pettini}, {Steidel}, {Rudie}  \&
  {Jorgenson}}{{Cooke} et~al.}{2011}]{Cooke2011}
{Cooke} R.,  {Pettini} M.,  {Steidel} C.~C.,  {Rudie} G.~C.,   {Jorgenson}
  R.~A.,  2011, \mn@doi [\mnras] {10.1111/j.1365-2966.2010.17966.x}, \href
  {https://ui.adsabs.harvard.edu/abs/2011MNRAS.412.1047C} {412, 1047}

\bibitem[\protect\citeauthoryear{{Dahn}, {Liebert}, {Kron}, {Spinrad}  \&
  {Hintzen}}{{Dahn} et~al.}{1977}]{Dahn1977}
{Dahn} C.~C.,  {Liebert} J.,  {Kron} R.~G.,  {Spinrad} H.,   {Hintzen} P.~M.,
  1977, \mn@doi [\apj] {10.1086/155518}, \href
  {https://ui.adsabs.harvard.edu/abs/1977ApJ...216..757D} {216, 757}

\bibitem[\protect\citeauthoryear{{\VAN{De}{De}{de} {Kool}} \&
  {Green}}{{\VAN{De}{De}{de} {Kool}} \& {Green}}{1995}]{deKool1995}
{\VAN{De}{De}{de} {Kool}} M.,  {Green} P.~J.,  1995, \mn@doi [\apj]
  {10.1086/176051}, \href
  {https://ui.adsabs.harvard.edu/abs/1995ApJ...449..236D} {449, 236}

\bibitem[\protect\citeauthoryear{{Dearborn}, {Liebert}, {Aaronson}, {Dahn},
  {Harrington}, {Mould}  \& {Greenstein}}{{Dearborn}
  et~al.}{1986}]{Dearborn1986}
{Dearborn} D.~S.~P.,  {Liebert} J.,  {Aaronson} M.,  {Dahn} C.~C.,
  {Harrington} R.,  {Mould} J.,   {Greenstein} J.~L.,  1986, \mn@doi [\apj]
  {10.1086/163805}, \href
  {https://ui.adsabs.harvard.edu/abs/1986ApJ...300..314D} {300, 314}

\bibitem[\protect\citeauthoryear{{Doi} et~al.,}{{Doi} et~al.}{2010}]{Doi2010}
{Doi} M.,  et~al., 2010, \mn@doi [\aj] {10.1088/0004-6256/139/4/1628}, \href
  {https://ui.adsabs.harvard.edu/abs/2010AJ....139.1628D} {139, 1628}

\bibitem[\protect\citeauthoryear{{Dotter}, {Chaboyer}, {Jevremovi{\'c}},
  {Kostov}, {Baron}  \& {Ferguson}}{{Dotter} et~al.}{2008}]{Dotter2008}
{Dotter} A.,  {Chaboyer} B.,  {Jevremovi{\'c}} D.,  {Kostov} V.,  {Baron} E.,
  {Ferguson} J.~W.,  2008, \mn@doi [\apjs] {10.1086/589654}, \href
  {https://ui.adsabs.harvard.edu/abs/2008ApJS..178...89D} {178, 89}

\bibitem[\protect\citeauthoryear{{Farihi}, {Arendt}, {Machado}  \&
  {Whitehouse}}{{Farihi} et~al.}{2018}]{Farihi2018}
{Farihi} J.,  {Arendt} A.~R.,  {Machado} H.~S.,   {Whitehouse} L.~J.,  2018,
  \mn@doi [\mnras] {10.1093/mnras/sty890}, \href
  {https://ui.adsabs.harvard.edu/abs/2018MNRAS.477.3801F} {477, 3801}

\bibitem[\protect\citeauthoryear{{Foreman-Mackey}, {Hogg}, {Lang}  \&
  {Goodman}}{{Foreman-Mackey} et~al.}{2013}]{Foreman-Mackey2013}
{Foreman-Mackey} D.,  {Hogg} D.~W.,  {Lang} D.,   {Goodman} J.,  2013, \mn@doi
  [\pasp] {10.1086/670067}, \href
  {https://ui.adsabs.harvard.edu/abs/2013PASP..125..306F} {125, 306}

\bibitem[\protect\citeauthoryear{{Frebel} \& {Norris}}{{Frebel} \&
  {Norris}}{2015}]{Frebel2015}
{Frebel} A.,  {Norris} J.~E.,  2015, \mn@doi [\araa]
  {10.1146/annurev-astro-082214-122423}, \href
  {https://ui.adsabs.harvard.edu/abs/2015ARA&A..53..631F} {53, 631}

\bibitem[\protect\citeauthoryear{{Fujimoto}, {Ikeda}  \& {Iben}}{{Fujimoto}
  et~al.}{2000}]{Fujimoto2000}
{Fujimoto} M.~Y.,  {Ikeda} Y.,   {Iben} Icko J.,  2000, \mn@doi [\apjl]
  {10.1086/312453}, \href
  {https://ui.adsabs.harvard.edu/abs/2000ApJ...529L..25F} {529, L25}

\bibitem[\protect\citeauthoryear{{Gaia Collaboration} et~al.,}{{Gaia
  Collaboration} et~al.}{2023}]{GaiaCollaboration2023}
{Gaia Collaboration} et~al., 2023, \mn@doi [\aap]
  {10.1051/0004-6361/202243940}, \href
  {https://ui.adsabs.harvard.edu/abs/2023A&A...674A...1G} {674, A1}

\bibitem[\protect\citeauthoryear{{Gass}, {Liebert}  \& {Wehrse}}{{Gass}
  et~al.}{1988}]{Gass1988}
{Gass} H.,  {Liebert} J.,   {Wehrse} R.,  1988, \aap, \href
  {https://ui.adsabs.harvard.edu/abs/1988A&A...189..194G} {189, 194}

\bibitem[\protect\citeauthoryear{{Ginsburg} et~al.,}{{Ginsburg}
  et~al.}{2019}]{astroquery}
{Ginsburg} A.,  et~al., 2019, \mn@doi [\aj] {10.3847/1538-3881/aafc33}, \href
  {https://ui.adsabs.harvard.edu/abs/2019AJ....157...98G} {157, 98}

\bibitem[\protect\citeauthoryear{{Green}}{{Green}}{2013}]{Green2013}
{Green} P.,  2013, \mn@doi [\apj] {10.1088/0004-637X/765/1/12}, \href
  {https://ui.adsabs.harvard.edu/abs/2013ApJ...765...12G} {765, 12}

\bibitem[\protect\citeauthoryear{{Green}}{{Green}}{2018}]{Green2018}
{Green} G.,  2018, \mn@doi [The Journal of Open Source Software]
  {10.21105/joss.00695}, \href
  {https://ui.adsabs.harvard.edu/abs/2018JOSS....3..695G} {3, 695}

\bibitem[\protect\citeauthoryear{{Green}, {Schlafly}, {Zucker}, {Speagle}  \&
  {Finkbeiner}}{{Green} et~al.}{2019}]{Green2019_ext}
{Green} G.~M.,  {Schlafly} E.,  {Zucker} C.,  {Speagle} J.~S.,   {Finkbeiner}
  D.,  2019, \mn@doi [\apj] {10.3847/1538-4357/ab5362}, \href
  {https://ui.adsabs.harvard.edu/abs/2019ApJ...887...93G} {887, 93}

\bibitem[\protect\citeauthoryear{{Gustafsson}, {Edvardsson}, {Eriksson},
  {J{\o}rgensen}, {Nordlund}  \& {Plez}}{{Gustafsson}
  et~al.}{2008}]{Gustafsson2008}
{Gustafsson} B.,  {Edvardsson} B.,  {Eriksson} K.,  {J{\o}rgensen} U.~G.,
  {Nordlund} {\r{A}}.,   {Plez} B.,  2008, \mn@doi [\aap]
  {10.1051/0004-6361:200809724}, \href
  {https://ui.adsabs.harvard.edu/abs/2008A&A...486..951G} {486, 951}

\bibitem[\protect\citeauthoryear{{Harris} et~al.,}{{Harris}
  et~al.}{1998}]{Harris1998}
{Harris} H.~C.,  et~al., 1998, \mn@doi [\apj] {10.1086/305908}, \href
  {https://ui.adsabs.harvard.edu/abs/1998ApJ...502..437H} {502, 437}

\bibitem[\protect\citeauthoryear{{Harris} et~al.,}{{Harris}
  et~al.}{2018}]{Harris2018}
{Harris} H.~C.,  et~al., 2018, \mn@doi [\aj] {10.3847/1538-3881/aac100}, \href
  {https://ui.adsabs.harvard.edu/abs/2018AJ....155..252H} {155, 252}

\bibitem[\protect\citeauthoryear{{Horta}, {Lu}, {Ness}, {Lisanti}  \&
  {Price-Whelan}}{{Horta} et~al.}{2024}]{Horta2024}
{Horta} D.,  {Lu} Y.~L.,  {Ness} M.~K.,  {Lisanti} M.,   {Price-Whelan} A.~M.,
  2024, \mn@doi [\apj] {10.3847/1538-4357/ad58de}, \href
  {https://ui.adsabs.harvard.edu/abs/2024ApJ...971..170H} {971, 170}

\bibitem[\protect\citeauthoryear{{Iben}}{{Iben}}{1965}]{Iben1965}
{Iben} Icko J.,  1965, \mn@doi [\apj] {10.1086/148429}, \href
  {https://ui.adsabs.harvard.edu/abs/1965ApJ...142.1447I} {142, 1447}

\bibitem[\protect\citeauthoryear{{Ivezi{\'c}} et~al.,}{{Ivezi{\'c}}
  et~al.}{2008}]{Ivezic2008}
{Ivezi{\'c}} {\v{Z}}.,  et~al., 2008, \mn@doi [\apj] {10.1086/589678}, \href
  {https://ui.adsabs.harvard.edu/abs/2008ApJ...684..287I} {684, 287}

\bibitem[\protect\citeauthoryear{{Jorissen} et~al.,}{{Jorissen}
  et~al.}{2016}]{Jorissen2016}
{Jorissen} A.,  et~al., 2016, \mn@doi [\aap] {10.1051/0004-6361/201526992},
  \href {https://ui.adsabs.harvard.edu/abs/2016A&A...586A.158J} {586, A158}

\bibitem[\protect\citeauthoryear{{Jura}, {Xu}  \& {Young}}{{Jura}
  et~al.}{2013}]{Jura2013}
{Jura} M.,  {Xu} S.,   {Young} E.~D.,  2013, \mn@doi [\apjl]
  {10.1088/2041-8205/775/2/L41}, \href
  {https://ui.adsabs.harvard.edu/abs/2013ApJ...775L..41J} {775, L41}

\bibitem[\protect\citeauthoryear{{Keller} et~al.,}{{Keller}
  et~al.}{2007}]{Keller2007}
{Keller} S.~C.,  et~al., 2007, \mn@doi [\pasa] {10.1071/AS07001}, \href
  {https://ui.adsabs.harvard.edu/abs/2007PASA...24....1K} {24, 1}

\bibitem[\protect\citeauthoryear{{Kilic}, {Munn}, {Harris}, {von Hippel},
  {Liebert}, {Williams}, {Jeffery}  \& {DeGennaro}}{{Kilic}
  et~al.}{2017}]{Kilic2017}
{Kilic} M.,  {Munn} J.~A.,  {Harris} H.~C.,  {von Hippel} T.,  {Liebert} J.~W.,
   {Williams} K.~A.,  {Jeffery} E.,   {DeGennaro} S.,  2017, \mn@doi [\apj]
  {10.3847/1538-4357/aa62a5}, \href
  {https://ui.adsabs.harvard.edu/abs/2017ApJ...837..162K} {837, 162}

\bibitem[\protect\citeauthoryear{{Kirkpatrick} et~al.,}{{Kirkpatrick}
  et~al.}{2024}]{Kirkpatrick2024}
{Kirkpatrick} J.~D.,  et~al., 2024, \mn@doi [\apjs] {10.3847/1538-4365/ad24e2},
  \href {https://ui.adsabs.harvard.edu/abs/2024ApJS..271...55K} {271, 55}

\bibitem[\protect\citeauthoryear{{Li} et~al.,}{{Li} et~al.}{2024}]{Li2024}
{Li} L.,  et~al., 2024, \mn@doi [\apjs] {10.3847/1538-4365/ad1881}, \href
  {https://ui.adsabs.harvard.edu/abs/2024ApJS..271...12L} {271, 12}

\bibitem[\protect\citeauthoryear{{Lindegren} et~al.,}{{Lindegren}
  et~al.}{2021}]{Lindegren2021}
{Lindegren} L.,  et~al., 2021, \mn@doi [\aap] {10.1051/0004-6361/202039653},
  \href {https://ui.adsabs.harvard.edu/abs/2021A&A...649A...4L} {649, A4}

\bibitem[\protect\citeauthoryear{{Lucatello}, {Tsangarides}, {Beers},
  {Carretta}, {Gratton}  \& {Ryan}}{{Lucatello} et~al.}{2005}]{Lucatello2005}
{Lucatello} S.,  {Tsangarides} S.,  {Beers} T.~C.,  {Carretta} E.,  {Gratton}
  R.~G.,   {Ryan} S.~G.,  2005, \mn@doi [\apj] {10.1086/428104}, \href
  {https://ui.adsabs.harvard.edu/abs/2005ApJ...625..825L} {625, 825}

\bibitem[\protect\citeauthoryear{{Luri} et~al.,}{{Luri}
  et~al.}{2018}]{Luri2018}
{Luri} X.,  et~al., 2018, \mn@doi [\aap] {10.1051/0004-6361/201832964}, \href
  {https://ui.adsabs.harvard.edu/abs/2018A&A...616A...9L} {616, A9}

\bibitem[\protect\citeauthoryear{{Majewski} et~al.,}{{Majewski}
  et~al.}{2017}]{Majewski2017}
{Majewski} S.~R.,  et~al., 2017, \mn@doi [\aj] {10.3847/1538-3881/aa784d},
  \href {https://ui.adsabs.harvard.edu/abs/2017AJ....154...94M} {154, 94}

\bibitem[\protect\citeauthoryear{{Margon}, {Kupfer}, {Burdge}, {Prince},
  {Kulkarni}  \& {Shupe}}{{Margon} et~al.}{2018}]{Margon2018}
{Margon} B.,  {Kupfer} T.,  {Burdge} K.,  {Prince} T.~A.,  {Kulkarni} S.~R.,
  {Shupe} D.~L.,  2018, \mn@doi [\apjl] {10.3847/2041-8213/aab42a}, \href
  {https://ui.adsabs.harvard.edu/abs/2018ApJ...856L...2M} {856, L2}

\bibitem[\protect\citeauthoryear{{Naidu}, {Conroy}, {Bonaca}, {Johnson},
  {Ting}, {Caldwell}, {Zaritsky}  \& {Cargile}}{{Naidu}
  et~al.}{2020}]{Naidu2020}
{Naidu} R.~P.,  {Conroy} C.,  {Bonaca} A.,  {Johnson} B.~D.,  {Ting} Y.-S.,
  {Caldwell} N.,  {Zaritsky} D.,   {Cargile} P.~A.,  2020, \mn@doi [\apj]
  {10.3847/1538-4357/abaef4}, \href
  {https://ui.adsabs.harvard.edu/abs/2020ApJ...901...48N} {901, 48}

\bibitem[\protect\citeauthoryear{{Plez} \& {Cohen}}{{Plez} \&
  {Cohen}}{2005}]{Plez2005}
{Plez} B.,  {Cohen} J.~G.,  2005, \mn@doi [\aap] {10.1051/0004-6361:20042082},
  \href {https://ui.adsabs.harvard.edu/abs/2005A&A...434.1117P} {434, 1117}

\bibitem[\protect\citeauthoryear{{Reid}, {Gizis}  \& {Hawley}}{{Reid}
  et~al.}{2002}]{Reid2002}
{Reid} I.~N.,  {Gizis} J.~E.,   {Hawley} S.~L.,  2002, \mn@doi [\aj]
  {10.1086/343777}, \href
  {https://ui.adsabs.harvard.edu/abs/2002AJ....124.2721R} {124, 2721}

\bibitem[\protect\citeauthoryear{{Riello} et~al.,}{{Riello}
  et~al.}{2021}]{Riello2021}
{Riello} M.,  et~al., 2021, \mn@doi [\aap] {10.1051/0004-6361/202039587}, \href
  {https://ui.adsabs.harvard.edu/abs/2021A&A...649A...3R} {649, A3}

\bibitem[\protect\citeauthoryear{{Roulston}, {Green}, {Toonen}  \&
  {Hermes}}{{Roulston} et~al.}{2021}]{Roulston2021}
{Roulston} B.~R.,  {Green} P.~J.,  {Toonen} S.,   {Hermes} J.~J.,  2021,
  \mn@doi [\apj] {10.3847/1538-4357/ac157c}, \href
  {https://ui.adsabs.harvard.edu/abs/2021ApJ...922...33R} {922, 33}

\bibitem[\protect\citeauthoryear{{Roulston} et~al.,}{{Roulston}
  et~al.}{2022}]{Roulston2022}
{Roulston} B.~R.,  et~al., 2022, \mn@doi [\apj] {10.3847/1538-4357/ac4706},
  \href {https://ui.adsabs.harvard.edu/abs/2022ApJ...926..210R} {926, 210}

\bibitem[\protect\citeauthoryear{{Salpeter}}{{Salpeter}}{1955}]{Salpeter1955}
{Salpeter} E.~E.,  1955, \mn@doi [\apj] {10.1086/145971}, \href
  {https://ui.adsabs.harvard.edu/abs/1955ApJ...121..161S} {121, 161}

\bibitem[\protect\citeauthoryear{{Sanders} \& {Binney}}{{Sanders} \&
  {Binney}}{2015}]{SandersBinney2015}
{Sanders} J.~L.,  {Binney} J.,  2015, \mn@doi [\mnras] {10.1093/mnras/stv578},
  \href {https://ui.adsabs.harvard.edu/abs/2015MNRAS.449.3479S} {449, 3479}

\bibitem[\protect\citeauthoryear{{Schlafly} \& {Finkbeiner}}{{Schlafly} \&
  {Finkbeiner}}{2011}]{SchlaflyFinkbeiner}
{Schlafly} E.~F.,  {Finkbeiner} D.~P.,  2011, \mn@doi [\apj]
  {10.1088/0004-637X/737/2/103}, \href
  {https://ui.adsabs.harvard.edu/abs/2011ApJ...737..103S} {737, 103}

\bibitem[\protect\citeauthoryear{{Starkenburg}, {Shetrone}, {McConnachie}  \&
  {Venn}}{{Starkenburg} et~al.}{2014}]{Starkenburg2014}
{Starkenburg} E.,  {Shetrone} M.~D.,  {McConnachie} A.~W.,   {Venn} K.~A.,
  2014, \mn@doi [\mnras] {10.1093/mnras/stu623}, \href
  {https://ui.adsabs.harvard.edu/abs/2014MNRAS.441.1217S} {441, 1217}

\bibitem[\protect\citeauthoryear{{Vasiliev}}{{Vasiliev}}{2019}]{Vasiliev2019}
{Vasiliev} E.,  2019, \mn@doi [\mnras] {10.1093/mnras/sty2672}, \href
  {https://ui.adsabs.harvard.edu/abs/2019MNRAS.482.1525V} {482, 1525}

\bibitem[\protect\citeauthoryear{{Venn}, {Irwin}, {Shetrone}, {Tout}, {Hill}
  \& {Tolstoy}}{{Venn} et~al.}{2004}]{Venn2004}
{Venn} K.~A.,  {Irwin} M.,  {Shetrone} M.~D.,  {Tout} C.~A.,  {Hill} V.,
  {Tolstoy} E.,  2004, \mn@doi [\aj] {10.1086/422734}, \href
  {https://ui.adsabs.harvard.edu/abs/2004AJ....128.1177V} {128, 1177}

\bibitem[\protect\citeauthoryear{{Whitehouse}, {Farihi}, {Green}, {Wilson}  \&
  {Subasavage}}{{Whitehouse} et~al.}{2018}]{Whitehouse2018}
{Whitehouse} L.~J.,  {Farihi} J.,  {Green} P.~J.,  {Wilson} T.~G.,
  {Subasavage} J.~P.,  2018, \mn@doi [\mnras] {10.1093/mnras/sty1622}, \href
  {https://ui.adsabs.harvard.edu/abs/2018MNRAS.479.3873W} {479, 3873}

\bibitem[\protect\citeauthoryear{{Whitehouse}, {Farihi}, {Howarth}, {Mancino},
  {Walters}, {Swan}, {Wilson}  \& {Guo}}{{Whitehouse}
  et~al.}{2021}]{Whitehouse2021}
{Whitehouse} L.~J.,  {Farihi} J.,  {Howarth} I.~D.,  {Mancino} S.,  {Walters}
  N.,  {Swan} A.,  {Wilson} T.~G.,   {Guo} J.,  2021, \mn@doi [\mnras]
  {10.1093/mnras/stab1913}, \href
  {https://ui.adsabs.harvard.edu/abs/2021MNRAS.506.4877W} {506, 4877}

\bibitem[\protect\citeauthoryear{{Yanny} et~al.,}{{Yanny}
  et~al.}{2009}]{Yanny2009}
{Yanny} B.,  et~al., 2009, \mn@doi [\aj] {10.1088/0004-6256/137/5/4377}, \href
  {https://ui.adsabs.harvard.edu/abs/2009AJ....137.4377Y} {137, 4377}

\bibitem[\protect\citeauthoryear{{Yong} et~al.,}{{Yong}
  et~al.}{2013}]{Yong2013}
{Yong} D.,  et~al., 2013, \mn@doi [\apj] {10.1088/0004-637X/762/1/27}, \href
  {https://ui.adsabs.harvard.edu/abs/2013ApJ...762...27Y} {762, 27}

\bibitem[\protect\citeauthoryear{{Yoon} et~al.,}{{Yoon}
  et~al.}{2016}]{Yoon2016}
{Yoon} J.,  et~al., 2016, \mn@doi [\apj] {10.3847/0004-637X/833/1/20}, \href
  {https://ui.adsabs.harvard.edu/abs/2016ApJ...833...20Y} {833, 20}

\bibitem[\protect\citeauthoryear{{York} et~al.,}{{York}
  et~al.}{2000}]{York2000}
{York} D.~G.,  et~al., 2000, \mn@doi [\aj] {10.1086/301513}, \href
  {https://ui.adsabs.harvard.edu/abs/2000AJ....120.1579Y} {120, 1579}

\bibitem[\protect\citeauthoryear{{Young}}{{Young}}{2014}]{Young2014}
{Young} E.~D.,  2014, \mn@doi [Earth and Planetary Science Letters]
  {10.1016/j.epsl.2014.02.014}, \href
  {https://ui.adsabs.harvard.edu/abs/2014E&PSL.392...16Y} {392, 16}

\bibitem[\protect\citeauthoryear{{Zepeda} et~al.,}{{Zepeda}
  et~al.}{2023}]{Zepeda2023}
{Zepeda} J.,  et~al., 2023, \mn@doi [\apj] {10.3847/1538-4357/dacbbcc}, \href
  {https://ui.adsabs.harvard.edu/abs/2023ApJ...947...23Z} {947, 23}

\bibitem[\protect\citeauthoryear{{Zhang}, {Iorio}, {Belokurov}, {Evans},
  {Bobrick}  \& {D'Orazi}}{{Zhang} et~al.}{2025}]{Zhang2025}
{Zhang} H.,  {Iorio} G.,  {Belokurov} V.,  {Evans} N.~W.,  {Bobrick} A.,
  {D'Orazi} V.,  2025, \mn@doi [arXiv e-prints] {10.48550/arXiv.2504.06720},
  \href {https://ui.adsabs.harvard.edu/abs/2025arXiv250406720Z} {p.
  arXiv:2504.06720}

\makeatother
\end{thebibliography}



\appendix

 \section{Data table}

The data for the high-confidence dC stars are provided in Table~\ref{tab:data}.

\begin{table}
    \caption{Description of the columns for the provided processed candidate dC stars. The columns \texttt{e\_X\_high} and \texttt{e\_X\_low} give the upper and lower $1\upsigma$ uncertainties of columns \texttt{X}. Energy, actions, and component probabilities are calculated using the model from \citet{Binney2023}. Halo probabilities can be computed as e.g.\ \texttt{probhalo}=1-\texttt{probthin}-\texttt{probthick}.}
    \centering
    \def\arraystretch{1.1}%
    \begin{tabular}{ll}
    Column & Description\\
    \hline
\texttt{name}&SDSS name from \citet{Green2013}\\
\texttt{source\_id}&{\it Gaia} DR3 source id\\
\texttt{ra}&{\it Gaia} DR3 right ascension [deg]\\
\texttt{dec}&{\it Gaia} DR3 declination [deg]\\
\texttt{angular\_separation}&\makecell[l]{Angular separation between SDSS DR16\\\quad and {\it Gaia} DR3 positions [arcsec]}\\
\texttt{M\_r}&\makecell[l]{Extinction-corrected $r$-band\\\quad absolute magnitude}\\
\texttt{g\_r}&de-reddened $(g-r)$ colour\\
\texttt{g\_i}&de-reddened $(g-i)$ colour \\
\texttt{r\_i}&de-reddened $(r-i)$ colour \\
\texttt{distance}&Distance [kpc]\\
\texttt{A\_r}&$r$-band extinction\\
\texttt{R}&Galactocentric radius [kpc]\\
\texttt{phi}&Galactocentric azimuth [rad]\\
\texttt{z}&Galactocentric vertical height [kpc]\\
\texttt{vR}&Galactocentric radial velocity [km\,s$^{-1}$]\\
\texttt{vphi}&Galactocentric azimuthal velocity [km\,s$^{-1}$]\\
\texttt{vz}&Galactocentric vertical velocity [km\,s$^{-1}$]\\
\texttt{energy}&Energy [(km\,s$^{-1}$)$^2$]\\
\texttt{JR}&Radial action [kpc km\,s$^{-1}$]\\
\texttt{Jz}&Vertical action [kpc km\,s$^{-1}$]\\
\texttt{Jphi}&\makecell[l]{$z$-component of angular\\\quad momentum [kpc km\,s$^{-1}$]}\\
\texttt{probthin}&Thin disc probability from ($\bm{x},\bm{v}$)\\
\texttt{probthick}&Thick disc probability from ($\bm{x},\bm{v}$)\\
\texttt{comp}&\makecell[l]{Most-likely component membership\\\quad (1=thin, 2=thick, 3=halo)}\\
\texttt{probthin\_x}&Thin disc probability from $\bm{x}$\\
\texttt{probthick\_x}&Thick disc probability from $\bm{x}$\\
\texttt{dC}&Dwarf carbon star boolean (True=dC)
    \end{tabular}
    \label{tab:data}
\end{table}

\section{Additional plots}\label{app:plots}

\begin{figure*}
\centering
\begin{subfigure}[b]{0.49\textwidth}
        \centering
        \includegraphics[width=\textwidth]{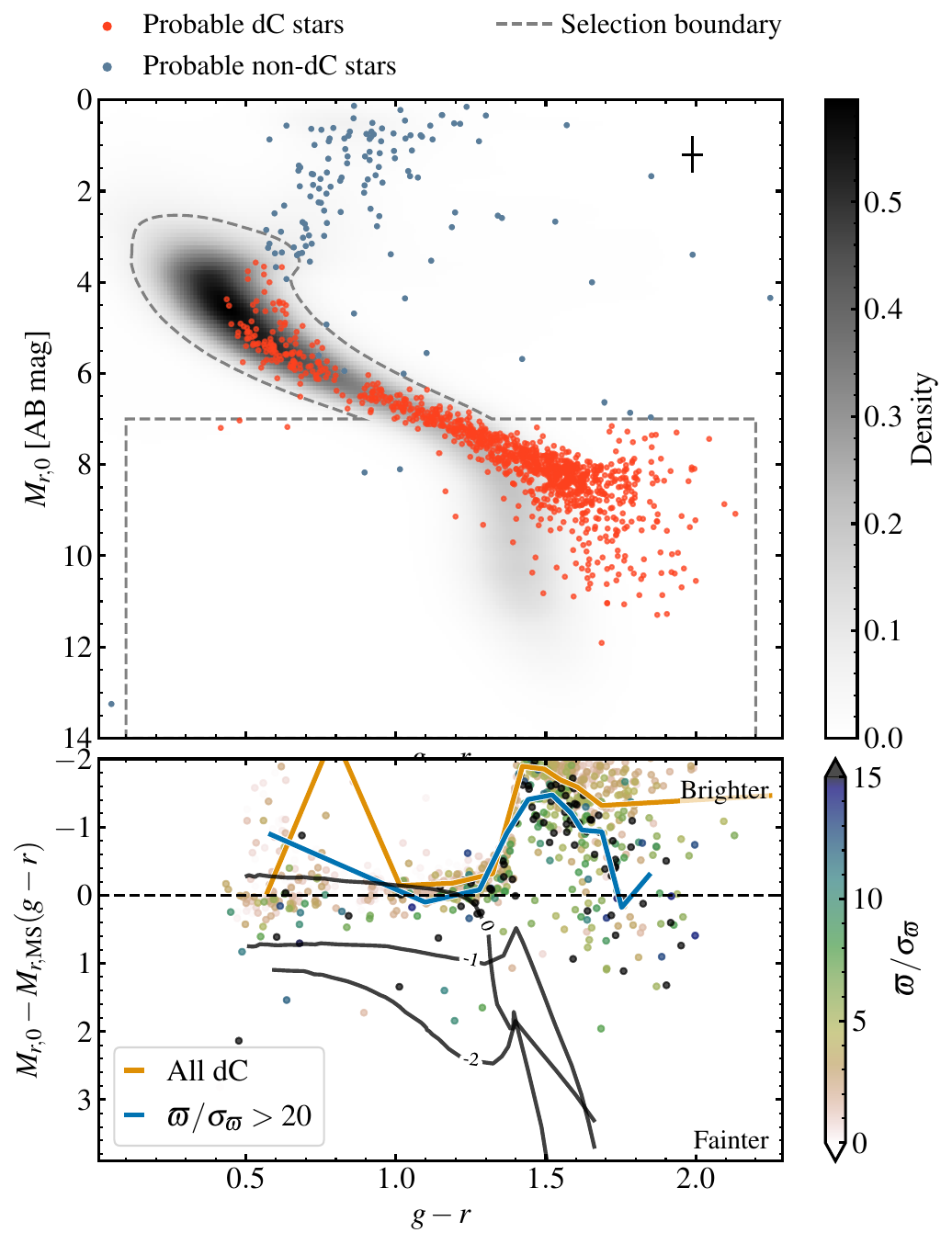} 
        \caption{$M_r$ vs.\ $(g-r)$ colour-magnitude diagram.}
        \label{fig:sub1}
    \end{subfigure}
    \hfill
    \begin{subfigure}[b]{0.49\textwidth}
        \centering
        \includegraphics[width=\textwidth]{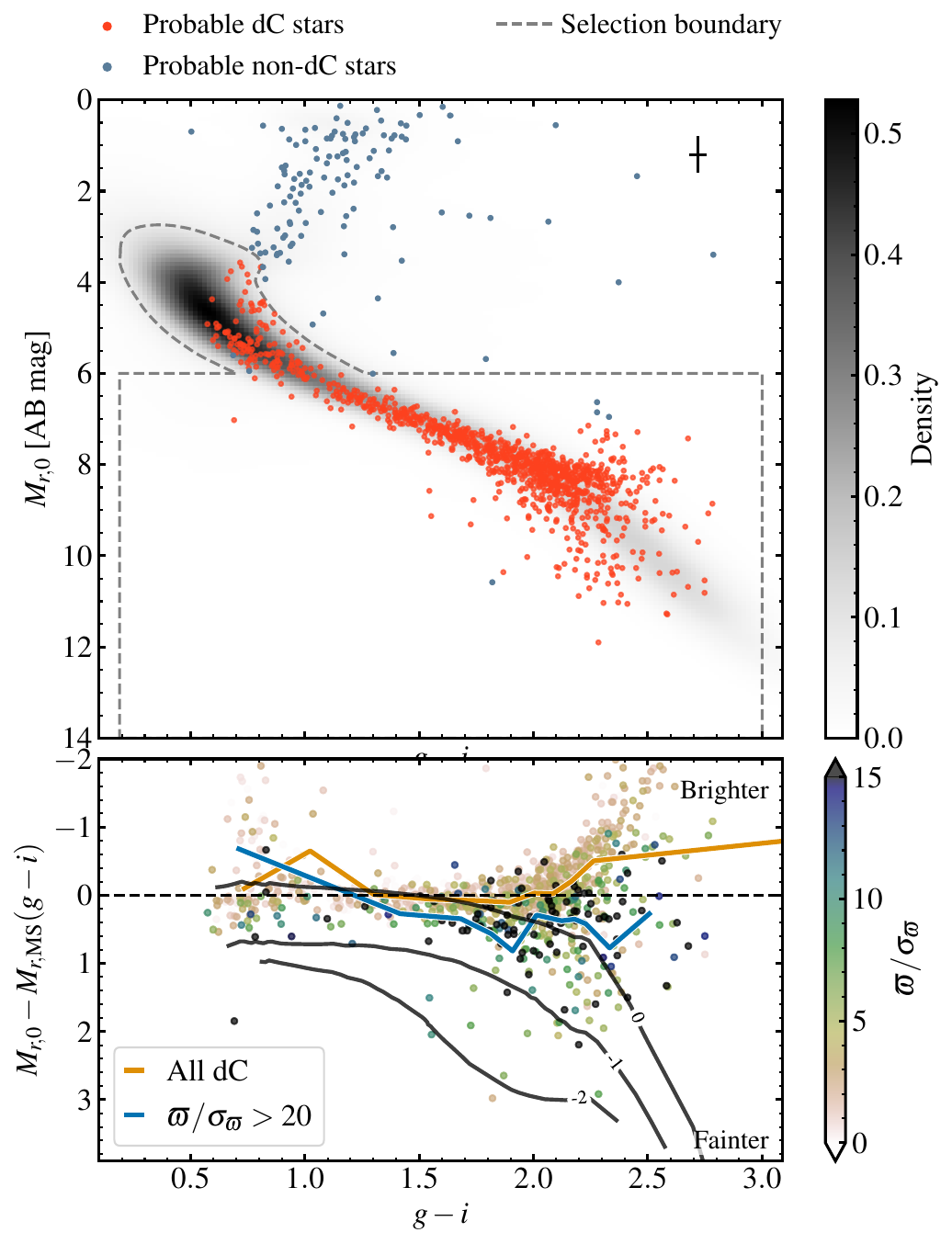}
        \caption{$M_r$ vs.\ $(g-i)$ colour-magnitude diagram.}
        \label{fig:sub2}
    \end{subfigure}
\caption{Additional colour-(absolute) magnitude diagrams for the candidate dC stars. See Figure~\ref{fig:CMD} for a description of the figures.}
\label{fig:extra_cmd}
\end{figure*}

This appendix provides additional supporting plots. Figure~\ref{fig:CMD} shows the $M_r$ vs.\ $(r-i)$ colour-magnitude diagram used for classifying high-confidence dC stars. Figure~\ref{fig:extra_cmd} shows $M_r$ vs.\ $(g-r)$ and $M_r$ vs.\ $(g-i)$ diagrams which have also been used as part of the classification. It is expected that the $g$ band photometry of dC stars is most distinct from that of carbon-normal stars owing to the presence of strong C$_2$ Swan absorption in this wavelength range, although the $r$ band is also moderately affected \citep[see figure~1 of][]{Green2013}. One additional interesting feature of these plots compared to Figure~\ref{fig:CMD} is that the dC stars with $\varpi/\upsigma_\varpi>20$ that are located near the main-sequence turn-off region are redder than the prior (and correspondingly the low parallax signal-to-noise dC candidates). Again, this is an indication of the carbon sensitivity of the $g$ band.

\begin{figure*}
    \centering
    \includegraphics[width=\textwidth]{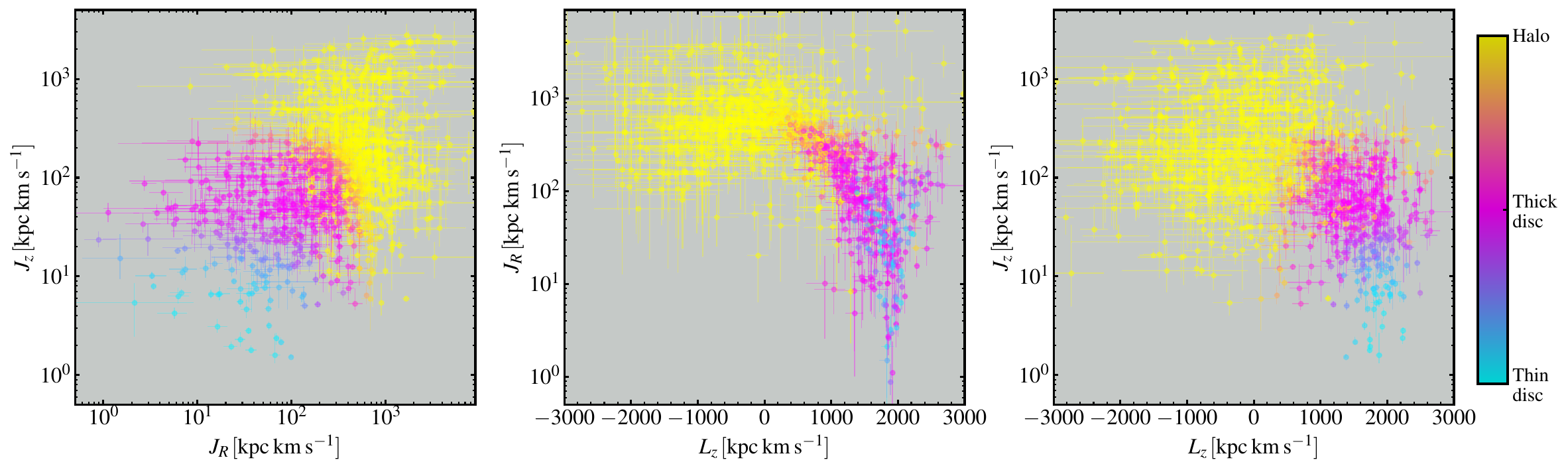}
\caption{Action distributions for the high-confidence dC stars. The points are coloured by their assigned component probability.}
    \label{fig:action_error}
\end{figure*}

Figure~\ref{fig:action_error} shows the action distribution of the high-confidence dC stars along with their uncertainties. The uncertainties can be substantial, particularly for the stars classified as halo, but sufficiently small to reliably distinguish the different components.

\bsp	
\label{lastpage}
\end{document}